\documentclass[prb,showpacs,superscriptaddress,twocolumn]{revtex4-1}
\usepackage{amssymb,amsmath,color,graphicx,psfrag}

\begin{document}

\newcommand{\kv}[0]{\mathbf{k}}
\newcommand{\Rv}[0]{\mathbf{R}}
\newcommand{\rv}[0]{\mathbf{r}}
\newcommand{\K}[0]{\mathbf{K}}
\newcommand{\Kp}[0]{\mathbf{K'}}
\newcommand{\dkv}[0]{\delta\kv}
\newcommand{\dkx}[0]{\delta k_{x}}
\newcommand{\dky}[0]{\delta k_{y}}
\newcommand{\dk}[0]{\delta k}
\newcommand{\cv}[0]{\mathbf{c}}
\newcommand{\qv}[0]{\mathbf{q}}
\newcommand{\pnt}[1]{\psfrag{#1}{\tiny{$#1$}}}

\newcommand{\jav}[1]{#1}

\title{Quantum quench in \jav{the Luttinger model} with finite temperature initial state}
\author{\'Ad\'am B\'acsi}
\email{bacsi@dept.phy.bme.hu}
\affiliation{Department of Physics, Budapest University of Technology and Economics, Budapest, Hungary}
\author{Bal\'azs D\'ora}
\affiliation{Department of Physics, Budapest University of Technology and Economics, Budapest, Hungary}
\affiliation{BME-MTA Exotic Quantum Phases Research Group, Budapest University of Technology and Economics, Budapest, Hungary}
\date{\today}

\begin{abstract}
We study the non-equilibrium dynamics of \jav{the Luttinger model} after a quantum quench, when the initial state is a finite temperature thermal equilibrium state. 
\jav{The  diagonal elements of the density matrix in the steady state show thermal features for high temperature initial states only, otherwise retain highly non-thermal character.
 The time evolution of Uhlmann fidelity, which measures the distance between the time evolved and initial states, is  evaluated for arbitrary initial temperatures and quench protocols. }
In the long time limit, the overlap between the time evolved and initial system decreases \jav{exponentially with the temperature with a universal prefactor.} 
Within perturbation theory, the statistics of final total energy and work are numerically evaluated in the case of a sudden quench\jav{, which yield identical distributions at zero temperature}. 
In both statistics, 
temperature effects are more significant in small systems.
\jav{The Dirac-delta peak at the adiabatic ground state energy remains present in the probability distribution of the total energy, but disappears
from the work distribution at non-zero initial temperatures.}

\end{abstract}


\maketitle

\section{Introduction}

Quantum quenches have been attracting lots of interest due to their experimentally controllable realizations in cold atomic systems \cite{greinernat,toshiyanat,hofferberthnat}. 
Recent experiments allow one to investigate quantum dynamics of closed systems and to perform quantum quenches by modulating the parameters of the system \cite{blochrmp,dziarmagareview}.
The quench drives the system out of equilibrium, raising interesting questions about how the closed system equilibrates after the quench, if at all, and how the long-time behaviour, i. e. the steady state, can be described.\cite{rigoltherm2,polkovnikovrmp,rigoltherm,rigoleth}

\jav{The characterization of the steady state can be given by determining all the diagonal density matrix elements in the representation of eigenstates of the final Hamiltonian. 
Off-diagonal elements do not contribute to the expectation value of physical observables in the steady state due to dephasing. This is the concept of the diagonal ensemble\cite{rigolgge,rigolnat,rigoltherm2,ggecaux}, 
which also enables us to determine the probability distribution of any constants of motion in the steady state.}


\jav{One-dimensional strongly correlated systems often form a Luttinger liquid (LL), made of bosonic sound-like collective excitations, regardless to the
statistics of the original system.
Such phases have already been created out of cold atoms\cite{hofferberthnat,cazalillarmp}.
However, it is not entirely clear whether the LL universality class can be extended to a nonequilibrium situation, though combined numerical and analytical  
studies indicate that this is indeed the case \cite{medenprl,balazslattnum,doraloschmidt}.
The non-equilibrium dynamics of the Luttinger model (LM), describing the low energy physics in LLs, has been studied extensively \cite{cazalillaprl,cazalillapra,balazsprl,perfetto,mitra,mitra1,nessiprb}.}
In our previous work, we have also investigated the statistics of work done on a LL -- prepared initially in the ground state -- during quantum quenches with different duration \cite{ztwork}. 

Finite temperature effects in thermalization\cite{rigolfintemp}, correlation functions and the momentum distribution function \cite{cazalillaprl,cazalillapra,tylutki} have already been investigated. In this paper, we study how the finite temperature modifies the time evolved density matrix and the diagonal ensemble if the {\jav{system}} is initially at thermal equilibrium, described by a canonical ensemble.

The time evolved state can be characterized by calculating the fidelity (or the Loschmidt echo) which measures the overlap with the initial thermal equilibrium state \cite{zanardipra,zanardi,damskiprl,zanardifintemp}. {\jav{The fidelity is an important quantity in various fields of physics ranging from nuclear physics to quantum information theory \cite{Gorin200633,goussevscholar} and also provides direct insight to dynamical properties of the quantum system without reference to any particular physical quantity.}}
In our model the final Hamiltonian does not commute with the initial density operator, therefore, the fidelity is 
expected to have explicit time-dependence {\jav{with equilibration in the steady state. How the time evolution and the long time behaviour depend on the initial temperature is one of the major concerns of the present work.}} 

With zero initial temperature, i. e., if the initial state is the pure ground state of the Hamiltonian, the statistics of total energy in the final state and the statistics of work done on the system are basically the same \cite{silvaqcp,ztwork}. At finite temperature, however, these distributions differ from each other because the initial energy is not well defined. Moreover, the initial energy may be arbitrarily large, therefore, the probability distribution function (PDF) of work has no lower bound\cite{hanggirmp,jarzynski}. 
In this paper our goal is to explore finite temperature effects in the statistics of total energy and work after a sudden quench (SQ).

The present article is organized as follows. After introducing the model and the time evolution during a quantum quench, we determine the diagonal ensemble for an arbitrary quench protocol and temperature in Section \ref{sec:de}. 
We derive exact analytical expressions for the Loschmidt echo in Section \ref{sec:fidmain}. The long time limit of the Loschmidt echo is numerically evaluated for SQs. In Section \ref{sec:tepdf} we study the statistics of the total energy in the SQ limit and within perturbation theory for weak interaction strength. The generating function of the distribution is obtained analytically while the PDFs are evaluated numerically with low initial temperature.
In Section \ref{sec:wpdf} the statistics of work is determined in the SQ limit.

\section{Time evolution during the quench}

{\jav{We study the time evolution of the LM described by the time-dependent bosonic Hamiltonian}}
\begin{gather}
\hat{H}(t)=\hat{H}_{0}+Q(t)\hat{V}, \label{hamilton}\\
\hat{H}_{0}=\sum_{q>0}\omega_{0}(q)\left(b^{+}_{q}b^{}_{q}+b^{+}_{-q}b^{}_{-q}\right),\\
\hat{V}=\sum_{q>0}\delta\omega(q)\left(b^{+}_{q}b^{}_{q}+b^{+}_{-q}b^{}_{-q}\right)+g(q)\left(b^{+}_{q}b^{+}_{-q}+b^{}_{q}b^{}_{-q}\right),
\end{gather} where
$\omega_{0}(q)=vq$ is the non-interacting dispersion of bosons, $\delta\omega(q)=\delta v q$ comes from velocity renormalization and $g(q)=g_{2}q\exp{(-R_{0}q)}$ is the interaction strength with $R_{0}$ characterizing the finite range of the interaction\cite{giamarchi,nersesyan}. 
In the following we use the 
notation $\omega(q,t)=\omega_{0}(q)+Q(t)\delta\omega(q)$ and $g(q,t)=Q(t)g(q)$. The quench protocol $Q(t)$ vanishes for $t<0$ and equals $1$ for $t>\tau$ with $\tau$ denoting the quench duration.

\jav{In equilibrium, the LM describes successfully the low energy dynamics of LLs. In a non-equilibrium situation, additional processes, which are termed irrelevant in equilibrium, are inevitably present in lattice models and
can still play an important
role.
To understand the applicability of the LM in non-equilibrium situation, several lattice models have been tested and investigated by comparing numerically exact calculations with analytical results using bosonization  
\cite{medenprl,balazslattnum,doraloschmidt,kennes}. These exhibit convincing agreement in all the examined cases.}

After the quench, the Hamiltonian $\hat{H}(\tau)=\hat{H}_{0}+\hat{V}$ can be diagonalized by standard Bogoliubov transformation, leading to 
\begin{gather}
\hat{H}(\tau)=E_{\mathrm{ad}}+\sum_{q>0}\Omega_{q}\left(d^{+}_{q}d_{q}+d^{+}_{-q}d_{-q}\right),
\end{gather}
where $\Omega(q)=\sqrt{\omega(q,\tau)^{2}-g(q)^{2}}<\omega(q,\tau)$ is the quasi-particle dispersion and $E_{\mathrm{ad}}=\sum_{q>0}\left(\Omega(q)-\omega(q,\tau)\right)<0$ is the ground state energy of the final Hamiltonian. The 
annihilation operators $b_{\pm q}$ is expressed with the new bosonic quasi-particle operators as
\begin{gather}
b_{\pm q}=\frac{\sqrt{\omega(q,\tau)+\Omega(q)}\,d_{\pm q}-\sqrt{\omega(q,\tau)-\Omega(q)}\,d_{\mp q}^{+}}{\sqrt{2\Omega(q)}}.
\end{gather}
We now focus on the time evolution of the density operator. The initial state is considered the finite temperature equilibrium state, $\hat{\rho}_{0}=\exp{(-\beta \hat {H}_{0})}/Z_{0}$, where $\beta=1/T$ is the inverse temperature.

The coupling between the system and reservoir is assumed to be so small, that the relaxation time of thermalization is much longer than the time-scale of the experiment. Energy exchange 
between the system and the environment is neglected apart from the energy change due to the quench.
Therefore, the time evolution is driven by the time dependent Schr\"odinger equation and can be transferred to the Bogoliubov coefficients defined through the time dependent 
creation and annihilation operators
\begin{equation}
\label{eq:btime}
b_{\pm q}(t)=u_{q}(t)b_{\pm q}+v_{q}^{*}(t)b_{\mp q}^{+}\,,
\end{equation}
where the $b$ bosons on the r.h.s. are those before the quench.
The coefficients are determined from Heisenberg's equation of motion and obey
\begin{equation}
\label{eq:tdschr}
i\partial_{t}\left[\begin{array}{c} u_{q}(t) \\ v_{q}(t) \end{array}\right]=\left[\begin{array}{cc} \omega(q,t) & g(q,t) \\ -g(q,t) & -\omega(q,t) \end{array}\right]\left[\begin{array}{c} u_{q}(t) \\ v_{q}(t) \end{array}\right]
\end{equation}
with the initial conditions $u_{q}(0)=1$ and $v_{q}(0)=0$ and  $|u_{q}(t)|^{2}-|v_{q}(t)|^{2}=1$. In the next section, we discuss the result of a general quench protocol on the diagonal ensemble, while in Sections IV and V we will focus on the SQ limit for the sake of 
simplicity. 

\section{Density operator and diagonal ensemble}

\label{sec:de}
The density matrix $\hat{\rho}(\tau)$ after the quench can be given exactly in second quantized formalism by means of $u_{q}(\tau)$ and $v_{q}(\tau)$. Using the exact expression (given in Appendix \ref{sec:Fxi}), 
we derive the diagonal elements of the density matrix which are essential to describe the steady state in the long time limit.

Since the various $q>0$ momentum modes are completely decoupled, as seen in Eq. 
\eqref{hamilton},  the density operator is block diagonal in momentum representation, and we consider only a single  $q>0$ channel in this section. 
The resulting Hamiltonian reduces to two coupled harmonic oscillators with the same frequency $\omega_{0}$.
Our results can easily be generalized to all channels by taking the product of the density operators of all modes. We also drop the subscript $q$ and the indices $q$ and $-q$ are replaced by $+$ and $-$, respectively (e.g. $d_{q}$ will henceforth be denoted with $d_{+}$).

After the quench the integrals of motion are $\hat{n}_{+}=d^{+}_{+}d_{+}$ and $\hat{n}_{-}=d^{+}_{-}d_{-}$. Of course, their products  and linear combinations are also preserved. The Hilbert space is spanned by the occupation number eigenstates $|n_{+},n_{-}\rangle$ in which the number of 
$d_{\pm}$  bosons is $n_{\pm}$.

In order to determine the diagonal ensemble, we calculate all the diagonal matrix elements
\begin{equation}
\label{eq:densme}
\rho(n_{+},n_{-}):=\langle n_{+},n_{-}|\hat{\rho}(\tau)|n_{+},n_{-}\rangle\,,
\end{equation}
which give the probability distribution of occupation numbers.

Off-diagonal elements, such as $\langle n'_{+},n'_{-}|\hat{\rho}(\tau)|n_{+},n_{-}\rangle$ which are non-zero only if $n_{+}-n_{-}=n'_{+}-n'_{-}$, are important only if the time evolution of a non-preserved quantity is studied. 
In the next section, for example, the fidelity will be such a quantity since $\hat{H}_{0}$ is not an integral of motion after the quench. In the long time limit, however, contributions from off-diagonal elements vanish due to dephasing which is a consequence of the continuous spectrum 
$\Omega(q)$\cite{barthel}.

The generating function of the occupation number distribution, Eq. \eqref{eq:densme} is obtained as
\begin{equation}
\label{eq:fdef}
f(\xi_{+},\xi_{-})=\mathrm{Tr}\,\left[\hat{\rho}(\tau)e^{i(\xi_{+}\hat{n}_{+}+\xi_{-}\hat{n}_{-})}\right]
\end{equation}
for all $\xi_{+}$ and $\xi_{-}$. The expectation value of all integrals of motion can be calculated by taking the derivatives of $f(\xi_{+},\xi_{-})$ with respect to $\xi_{+}$ and $\xi_{-}$. For instance, $\langle 
\hat{n}_{+}\rangle=-i\partial_{\xi_{+}}\left.f(\xi_{+},\xi_{-})\right|_{\xi_{+}=0,\xi_{-}=0}$ in 
the steady state. We will see in 
Section \ref{sec:tepdf} that Eq. \eqref{eq:fdef} is very useful in determining the generating function of the probability distribution of the total energy as well. It is important to realize that both the expectation value and the whole distribution of any integrals of motion can be deduced 
from 
$f(\xi_{+},\xi_{-})$.

\begin{figure}[t!]
\centering
a)\, \includegraphics[scale=0.22]{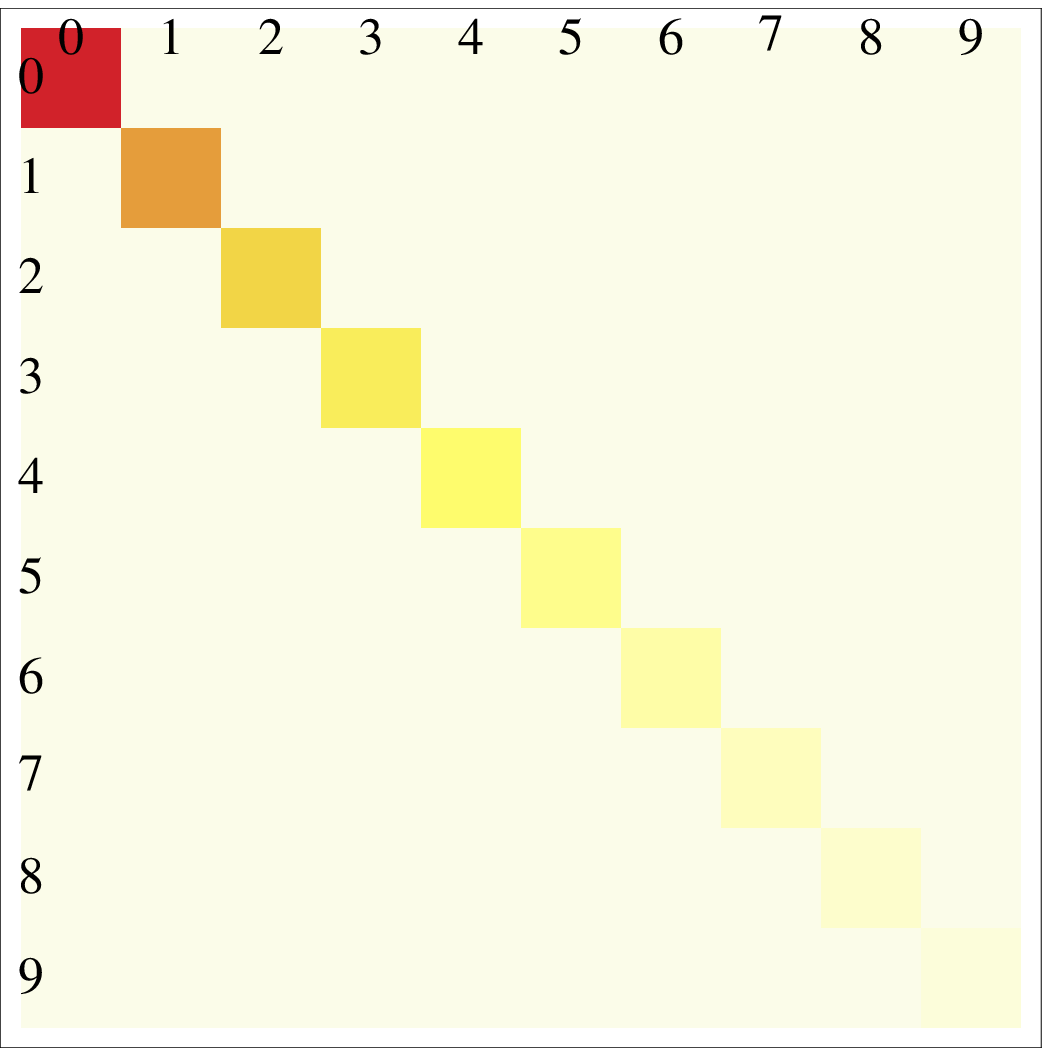}\,
b)\, \includegraphics[scale=0.22]{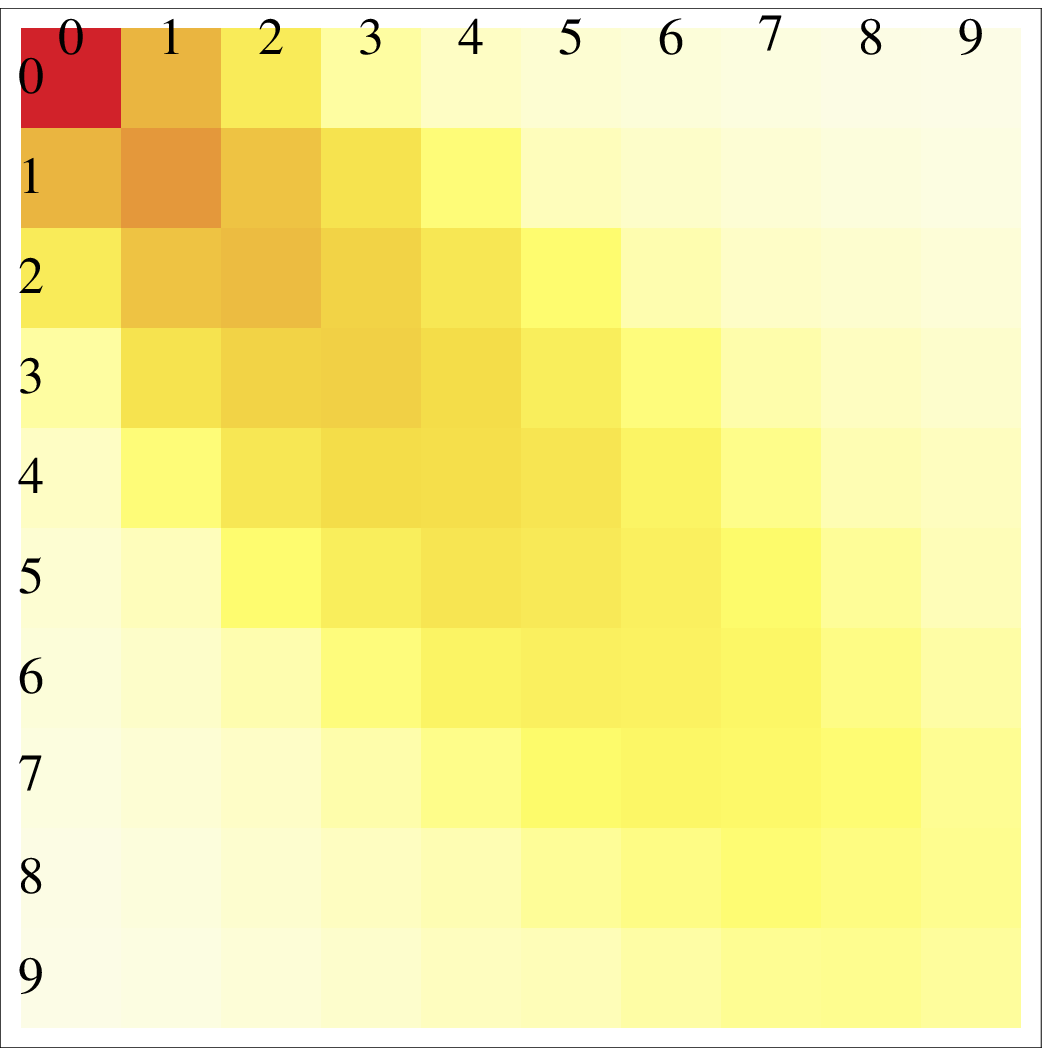}\, 
c)\,\includegraphics[scale=0.22]{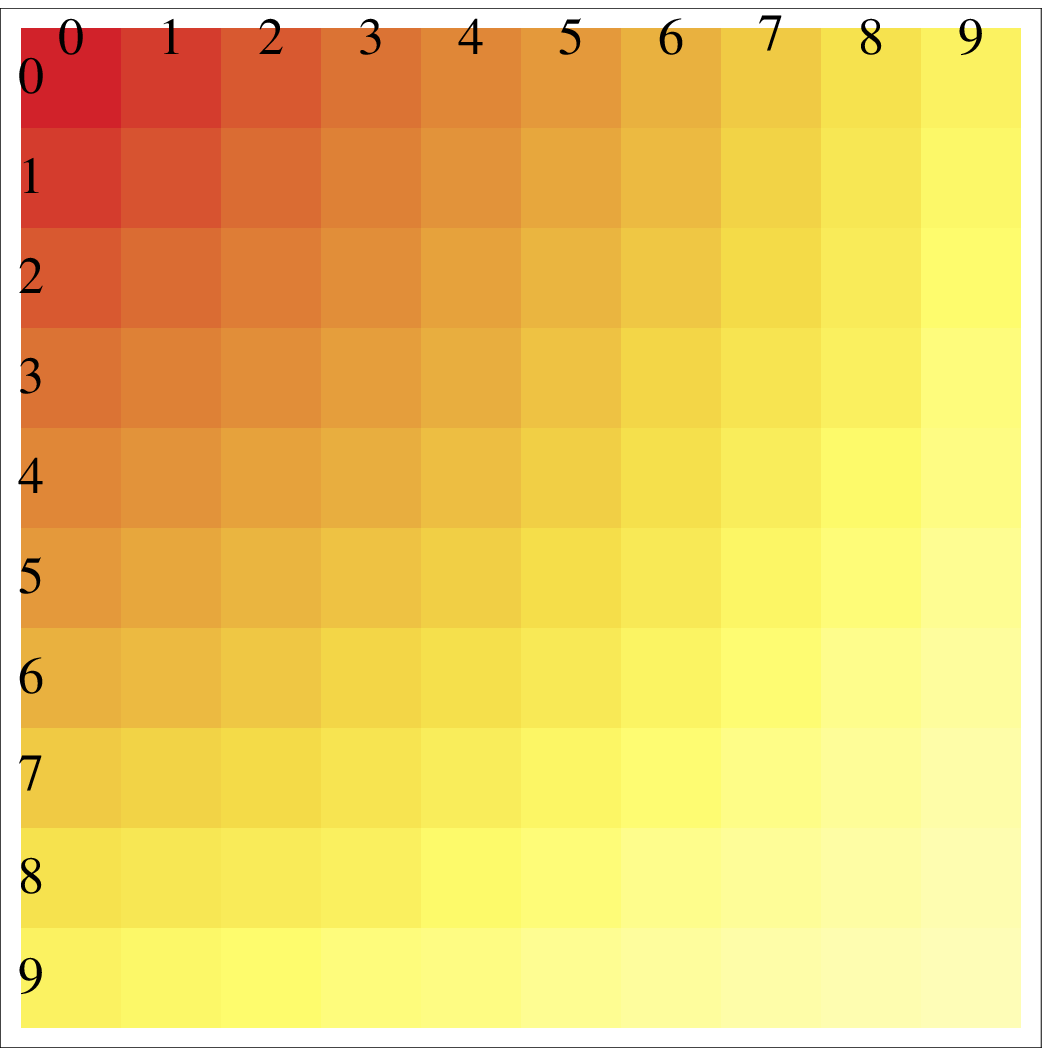}
\caption{Diagonal matrix elements of the density operator $\langle n_{+},n_{-}|\hat{\rho}(\tau)|n_{+},n_{-}\rangle$. In the figures $n_{+}$ and $n_{-}$ are measured on the horizontal and the vertical axes. The quantity $a(\tau)$, which characterizes the quench protocol and is independent from the temperature, is $2$ for all figures. The temperature varies such that the initial occupation number is a) $n_{0}=0$ zero temperature b) $n_{0}=1$ c) $n_{0}=30$ high temperature. Colors do not represent the same values in different figures and only illustrate the structure of diagonal matrix elements.}
\label{fig:denm}
\end{figure}

The function $f(\xi_{+},\xi_{-})$ is obtained for an arbitrary quench protocol and arbitrary temperature analytically. The detailed calculation is given in Appendix A and results in
\begin{gather}
f(\xi_{+},\xi_{-})=\Big[1+n(\tau)\left(1-e^{i(\xi_{+}+\xi_{-})}\right)+\nonumber \\
\label{eq:Fpm}
+\left(n_{0}+n_{0}^{2}\right)\left(e^{i\xi_{+}}-1\right)\left(e^{i\xi_{-}}-1\right)\Big]^{-1}
\end{gather}
where $n_{0}=(e^{\beta\omega_{0}}-1)^{-1}$ is the expectation value of the occupation number in the initial state and
\begin{gather}
n(\tau):=\mathrm{Tr}\left[\hat{\rho}(\tau)\hat{n}_{\pm}\right]=a(\tau)n_{0}+\frac{a(\tau)-1}{2}
\end{gather}
is the expectation value of the occupation number after the quench\cite{tylutki}. The real, \jav{temperature independent} quantity
\begin{equation}
\label{eq:atau}
a(\tau)=\frac{\omega(\tau)}{\Omega}\left(1+2|v(\tau)|^{2}\right)+2\frac{g}{\Omega}\textmd{Re}\left(u(\tau)^{*}v(\tau)\right)
\end{equation}
characterizes the quench protocol and does not depend on time for $t>\tau$ since it is related to $n(\tau)$, being the expectation value of the preserved quantity $\hat{n}_{\pm}$. \jav{We note that, since $a(\tau)\geq 1$, the average occupation number after the quench is larger than before, i.e. during the time evolution more bosons are created than annihilated on average. Moreover, the difference $n(\tau)-n_{0}$ grows as the initial temperature increases.}

The diagonal matrix elements are obtained by Fourier transforming Eq. \eqref{eq:Fpm}. This is carried out analytically by introducing the complex variables $z_{\pm}=e^{-i\xi_{\pm}}$. Then, complex integrals provide the matrix elements
\begin{eqnarray}
\label{eq:temp}
\rho(n_{+},n_{-})=\frac{(n_{0}+n_{0}^{2})^{n_{+}-n_{-}}\left(n(\tau)-n_{0}-n_{0}^{2}\right)^{n_{-}}}{\left(1+n(\tau)+n_{0}+n_{0}^{2}\right)^{n_{+}+1}}\nonumber\\
\times\sum_{l=0}^{n_{-}}\frac{\left(\begin{array}{c} n_{-} \\ l \end{array}\right)\left(\begin{array}{c} n_{+}+l \\ n_{-} \end{array}\right)\left(n_{0}+n_{0}^{2}\right)^{2l}}
{\left(n(\tau)-n_{0}-n_{0}^{2}\right)^{l}\left(1+n(\tau)+n_{0}+n_{0}^{2}\right)^{l}}
\end{eqnarray}
for $n_{+}\geq n_{-}$. The opposite case is obtained from $\rho(n_{+},n_{-})=\rho(n_{-},n_{+})$. {\jav{We emphasize again that Eq. \eqref{eq:temp} is exact for arbitrary quench protocol and initial temperature. All information about the quench is incorporated into the expectation value of the occupation number $n(\tau)$.}}

\jav{Let's start to analyze our results at zero temperatures first.} Only the $l=0$ term is finite in the sum in Eq. \eqref{eq:temp} and only the $n_{+}=n_{-}$ matrix element survives, 
meaning that the number of bosons in the $+q$ and $-q$ modes are the same. This behaviour stems from the fact that the difference 
$\hat{n}_{+}-\hat{n}_{-}$ is preserved during the time evolution and its 
expectation value is zero in the initial state\cite{ztwork}. \jav{The density matrix is highly non-thermal.}

In the finite temperature initial state, however, this difference may be nonzero and, therefore, $\rho(n_{+},n_{-}\neq n_{+})$ elements show up in the final state. 
Diagonal elements of the density matrix are illustrated in Fig. \ref{fig:denm} with different initial temperatures.

At low temperature or for high frequencies ($\beta\omega_{0}\gg 1$) the occupation number $n_{0}$ is exponentially small, $n_{0}\approx e^{-\beta\omega_{0}}$ to leading order. 
Matrix elements up to first order in $n_{0}$ are non-vanishing only if $n_{+}=n_{-}$ or $n_{+}=n_{-}\pm 1$, \jav{and the corrections to the zero temperature case are given by}
\begin{gather}
\rho(n_{+},n_{-})=\frac{2(a(\tau)-1)^{n_{+}}}{(a(\tau)+1)^{n_{+}+1}}
\begin{cases} 1-2n_{0} & \mathrm{if}\, n_{+}=n_{-} \\ {\displaystyle\frac{2n_{0} n_{\pm}}{a(\tau)\mp 1}} & \mathrm{if}\,\, n_{+}=n_{-}\pm 1 \end{cases},
\end{gather}
which is highly non-thermal again.

\jav{Finally,} at high temperature or for low frequencies ($\beta\omega_{0}\ll 1$) the initial occupation number is large ($n_{0}\gg 1$). If $n_{0}\gg a(\tau)$ also holds, the elements of the density matrix are written as
\begin{gather}
\rho(n_{+},n_{-})\approx\frac{1}{n_{0}^{2}} \exp{\left(-(n_{+}+n_{-})\frac{a(\tau)}{n_{0}}\right)},
\end{gather}
which resemble to a thermal density matrix. Note that the trace of this approximate matrix does not yield $1$ due to the high temperature approximation.

We emphasize again that only a single $q>0$ mode was considered in this section. 
All the modes should be taken into account when certain physical quantities, e.g. total energy, are evaluated.

\section{Loschmidt echo}
\label{sec:fidmain}

Here we investigate the question of how much the time evolved state described by $\hat{\rho}(t)$ differs from the initial state $\hat{\rho}_{0}$ for $t>\tau$. The physical quantity measuring the "similarity" of these states, i.e. the overlap of the two density operators, is the fidelity or sometimes called Loschmidt echo \cite{zanardi,zanardifintemp}. 
Since the initial Hamiltonian is not a constant of motion after the quench, the fidelity has explicit time-dependence \cite{aperes}. 

\jav{The distinguishability of quantum states is measured by means of the so-called Uhlmann fidelity \cite{jozsauhlmann,zanardi,nielsen}, which 
is defined as
\begin{gather}
F_{U}(t)=\mathrm{Tr}\left[\sqrt{\hat{\rho}^{1/2}_{0} \hat{\rho}(t)\hat{\rho}^{1/2}_{0}}\right]\,.
\label{uhlmann}
\end{gather}
The fidelity is symmetric with respect to its arguments and $0\leq F_{U}(t)\leq 1$ always holds where the latter relation becomes equality in the case of identical density operators. 
The Uhlmann fidelity is related to the Bures metric in which the angle between  the two density matrices is given by the angle $\arccos F_{U}(t)$. Since the density matrices are normalized to unity, their angle can be used to
quantify their distance.
In the case of pure states, the Uhlmann fidelity simplifies to the absolute value of the overlap between the wavefunctions.
For instance, with zero initial temperature the fidelity yields $F_{U}(t)=|\langle\Psi(t)|\Psi_{0}\rangle|$ where $|\Psi_{0}\rangle$ is the ground state of the initial Hamiltonian and $|\Psi(t)\rangle$ is the time evolved wavefunction. 
At finite initial temperature, evaluation of the trace provides
\begin{equation}
\label{eq:uhlmannfid}
\ln F_{U}(t)=\sum_{q>0}\ln\frac{\cosh(\beta\omega_{0}(q))-1}{\sqrt{1+|u_{q}(t)|^{2}\sinh^{2}(\beta\omega_{0}(q))}-1}
\end{equation}
where $u_{q}(t)$ is the Bogoliubov coefficient defined in Eq. \eqref{eq:btime}. For technical details, see Appendix \ref{sec:fid}. The resulting expression in Eq. \eqref{eq:uhlmannfid} shows that the Loschmidt echo depends remarkably on the initial temperature. 
This property seems to be natural but if we choose the Frobenius norm instead of the Bures metric, no initial temperature dependence is found. }

\jav{Let us briefly mention that Eq. \eqref{uhlmann}
represents the trace norm of the operator $\hat{\rho}^{1/2}_{0} \hat{\rho}^{1/2}(t)$. However, one can also use the Frobenius norm \cite{zanardi} instead, 
which also coincides with the finite temperature generalization of the Loschmidt echo as given in the pioneering paper by A. Peres\cite{aperes}.
In this case the overlap of the time-evolved and initial states is given by
\begin{gather}
\label{eq:ffdef}
F_{F}(t)=\sqrt{\mathrm{Tr}\left[\hat{\rho}(t)\hat{\rho}_{0}\right]}
\end{gather}
and  
\begin{equation}\label{eq:frobfid}\ln F_{F}(t)=\sum_{q>0}\ln\frac{\cosh(\beta\omega_{0}(q))-1}{|u_{q}(t)|\sinh(\beta\omega_{0}(q))}\,.\end{equation} 
Eq. \eqref{eq:ffdef} does not necessarily yield $1$ in the case of identical operators. 
Therefore, the Loschmidt echo is normalized by the square root of the so called effective dimension\cite{zanardi}  $d_{\mathrm{eff}}=1/\mathrm{Tr}[\rho_{0}^{2}]$. 
This normalization leads to
\begin{equation}
\label{eq:ffnorm}
\ln\big(\sqrt{d_{\mathrm{eff}}}F_{F}(t)\big)=-\sum_{q>0}\ln\left|u_{q}(t)\right|
\end{equation}
which leads to the rather counterintuitive result that the Loschmidt echo using the Frobenius norm does not depend on the initial temperature. 
This means that the Uhlmann fidelity, as used in quantum information theory, enables us to distinguish the time-evolved and initial states with finite temperature in a more delicate way than using the Frobenius norm.}

Eqs. \eqref{eq:uhlmannfid}, \eqref{eq:frobfid} and \eqref{eq:ffnorm} fulfill the inequality 
\begin{gather}
F_{F}(t)\leq F_{U}(t)\leq \sqrt{d_{\mathrm{eff}}}F_{F}(t),
\end{gather}
where the first relations holds true in general, while the second inequality is specific to bosonic systems, and is reversed for fermions \cite{zanardi}. The zero temperature limit of the Uhlmann fidelity yields the normalized Frobenius fidelity.

Eqs. \eqref{eq:frobfid}, \eqref{eq:ffnorm} and \eqref{eq:uhlmannfid} are the main results for the finite temperature Loschmidt echo, 
valid for arbitrary temperature, quench protocol and interaction strength for quadratic bosonic Hamiltonians. Previous studies of the fidelity at finite temperatures focused on 
fermionic 
systems\cite{zanardi}, 
though
the fidelity susceptibility for bosons was also considered\cite{sirker}.

\begin{figure}[h!]
\centering
\psfrag{Fxaxis}{$g_{2}/v$}
\psfrag{hh}{higher $T$}
\psfrag{FyaxisFyaxis}{$\ln F_{U}(t\rightarrow\infty)$}
\psfrag{time}{$t/\tau_{0}$}
\psfrag{fidfidfid}{$\ln F_{U}(t)$}
\pnt{-}
\pnt{0}\pnt{2}\pnt{4}\pnt{6}\pnt{8}\pnt{10}\pnt{12}\pnt{14}\pnt{0.00}\pnt{0.02}\pnt{0.04}\pnt{0.06}\pnt{0.08}\pnt{0.10}\pnt{0.12}
\pnt{0.0}\pnt{0.1}\pnt{0.2}\pnt{0.3}\pnt{0.4}\pnt{0.6}\pnt{0.8}\pnt{1.0}
a)\quad\includegraphics[scale=0.6]{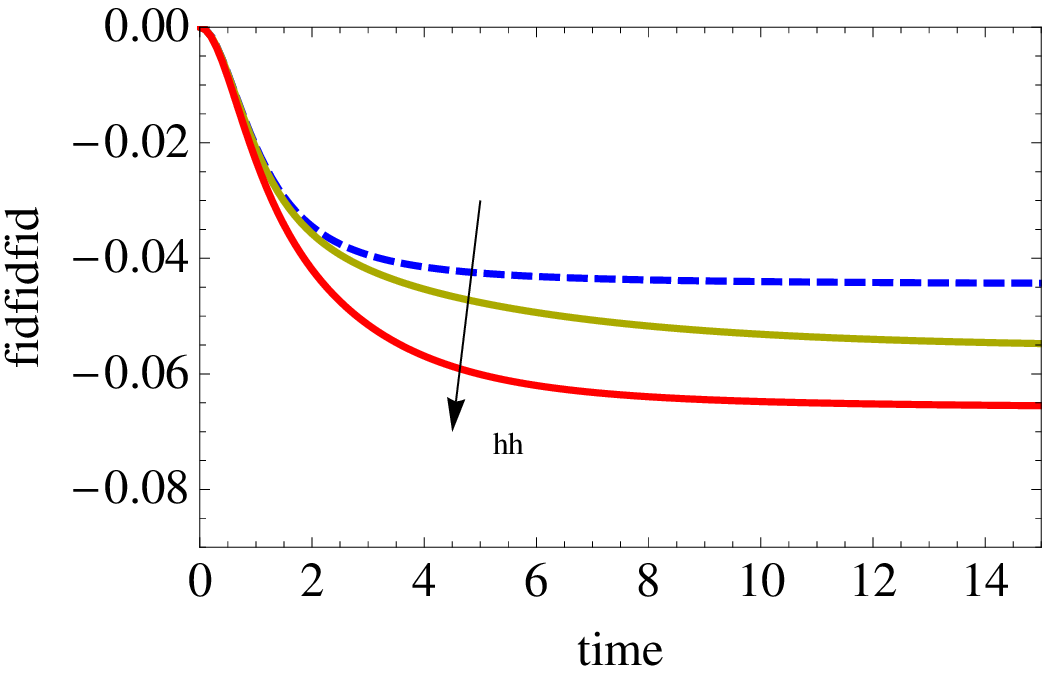}\\\quad\\
b)\quad\includegraphics[scale=0.6]{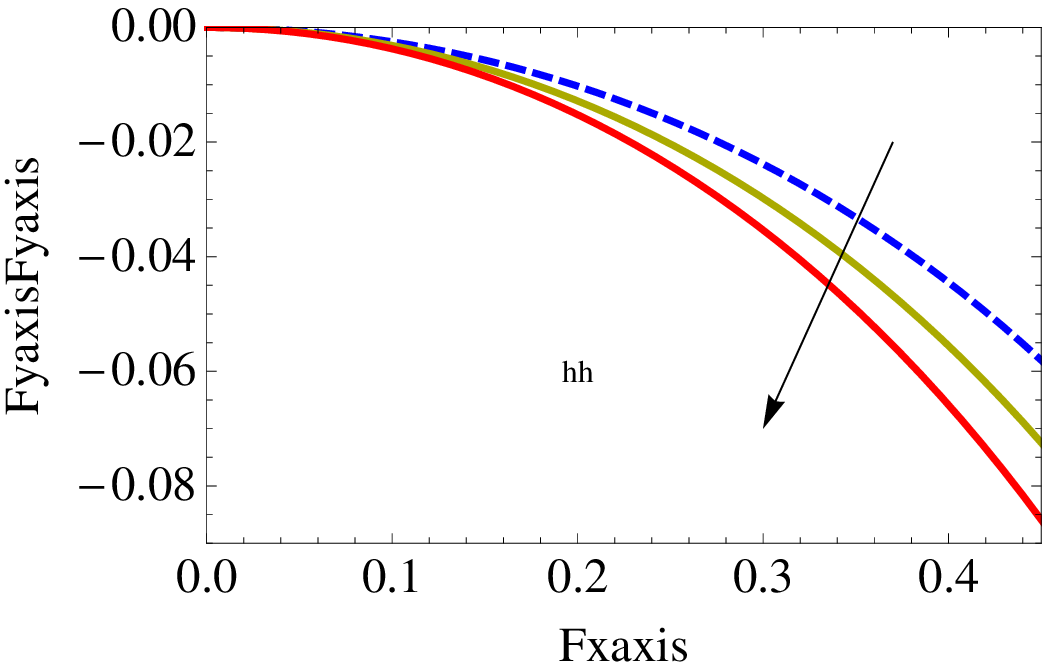}
\caption{a) Time evolution of the logarithm of the Loschmidt echo using the Bures metric for different temperatures following a sudden quench in units of $L/(4\pi R_{0})$. The interaction strength was chosen $g_{2}=0.4 v$ for the plot. b) Logarithm of the long time limit of the Loschmidt echo, in units of $L/(4\pi R_{0})$,  as a function of $g_{2}/v$  
(numerical results).
In both figures, the blue dashed line corresponds to the zero temperature case. The yellow and red curves correspond to finite temperatures $\beta=10\tau_{0}$ and $\beta=4\tau_{0}$, respectively, where $\tau_{0}=R_{0}/v$ is related to the finite range of the interaction.}
\label{fig:fid}
\end{figure}

In the following of this section, we investigate the Uhlmann fidelity in special cases.
In the SQ limit ($\tau\rightarrow 0$), the Bogoliubov coefficient is obtained as 
\begin{gather}
u_{q}(t)=\cos(\Omega(q)t)-i\frac{\omega(q,\tau)}{\Omega(q)}\sin(\Omega(q)t)
\end{gather}
for $t>0$. 
The time dependence of the Uhlmann fidelity is evaluated numerically, and the resulting function decreases monotonically but saturates to a non-zero value as shown in Fig. \ref{fig:fid}.a. 
Here, the velocity renormalization is neglected in this calculation because we are interested in interaction effects coming from a finite $g_{2}$.
Figure \ref{fig:fid}.b shows the long time limit of the Uhlmann fidelity as a function of $g_{2}/v$.
The final state deviates from the initial one with increasing temperature in the long time limit.

\jav{Analytical results are obtained only within perturbation theory for small values of $g_{2}/v$, when Eq. \eqref{eq:uhlmannfid} is expanded  in $|v_q(t)|$. 
At low temperatures and for a SQ, its long time value ($t\gg\beta\gg\tau_{0}$) is obtained as
\begin{gather}
\label{eq:fusq}
F_{U}(t\rightarrow\infty)=\exp\left(-\alpha-\frac{1}{16}\left(\frac{g_2}{v}\right)^2\frac{LT}{v}\right),
\end{gather}
where $\tau_{0}=R_{0}/v$ is the time scale corresponding to the finite range of the interaction, and the temperature dependent term possesses a universal prefactor in the exponent in the LL sense, namely
that it is independent of the high energy cutoff, $1/\tau_0$.
This universality is reminiscent of the universal term in the partition function and consequently in the specific heat of 1D critical quantum systems\cite{giamarchi}. 
Similar behaviour of the fidelity was reported in Ref. \onlinecite{sirker},  where the finite temperature fidelity susceptibility was investigated between different LM ground states. 
Our result also shows that increasing temperature results in less fidelity. This behaviour is related to the fact that more bosons are created during the quench for higher initial temperatures. In Eq. \eqref{eq:fusq}},  $\alpha=|E_{\mathrm{ad}}|\tau_{0}$ is the 
orthogonality exponent with $E_{\mathrm{ad}}=-Lg_{2}^{2}/(16\pi\tau_{0}^{2}v^{3})$ being the ground state energy of the final Hamiltonian within perturbation theory, $L$ is the length 
of the sample\cite{ztwork}.
\jav{In Ref. \onlinecite{doraloschmidt}, it was shown that the long time limit of the zero temperature
Loschmidt echo obtained from the LM describes that of the XXZ Heisenberg spin chain.
We believe that this agreement can be extended to finite temperatures, given the fact, that
 finite temperatures mostly affect states with energy smaller than $T$. Therefore, as long as $T\ll 1/\tau_0$, these corrections are expected to be universal, as demonstrated in Eq. \eqref{eq:fusq}.}
Even in the long time limit, we require $t\ll L/v$. \jav{For larger times, comparable to $L/v$, quantum revival occurs similarly to other cases\cite{zanardi}, which is beyond the scope of the present paper.}

\section{Probability distribution of total energy after quantum quench}
\label{sec:tepdf}

In this section we analyze the statistics of the total final energy, which, as opposed to work statistics, requires only one energy measurement. Therefore, repeating the procedure of releasing the LL from the trap and measuring its energy many times is expected to lead
to the probability distribution function of the total energy. 

\jav{Since the total energy is preserved after the quench, its distribution does not change while the steady state is reached. Therefore, it is sufficient to determine the PDF right at $t=\tau$.}
Its generating function is defined as
\begin{equation}\label{eq:gdef}G(\lambda)=\mathrm{Tr}\left[\hat{\rho}(\tau)e^{i\lambda \hat{H}(\tau)}\right]\,. 
\end{equation}
Since the Hamiltonian $\hat{H}(\tau)$ is a linear combination of $\hat{n}_{q}$ and $\hat{n}_{-q}$, the generating function can be given by means of $f(\xi_{+},\xi_{-})$ defined in Eq. \eqref{eq:fdef}. From this, we get
\begin{widetext}
\begin{equation}
\label{eq:lngfull}
\ln G(\lambda)=i\lambda E_{\mathrm{ad}}+\sum_{q>0}\ln f(\lambda\Omega(q),\lambda\Omega(q))=
i\lambda E_{\mathrm{ad}}-\sum_{q>0}\ln\left[1+n(\tau,q)\left(1-e^{2i\lambda\Omega(q)}\right)-\frac{\partial n_{0}(q)}{\partial(\beta\omega_{0}(q))}\left(e^{i\lambda\Omega(q)}-1\right)^{2}\right]
\end{equation}
\end{widetext}
where $n(\tau,q)$ and $n_{0}(q)$ are the expectation value of the occupation number after and before the quench, respectively.
The summation over $q$ cannot be performed analytically, therefore to make progress, we assume a small  $g_{2}/v$ and consider the generating function within perturbation theory for the case of SQ. We also disregard the velocity renormalization.
 The occupation number in the final state is \begin{gather}n(\tau=0,q)=n_{0}(q)+\frac{g(q)^{2}}{4\omega_{0}(q)^{2}}\left(n_{0}(q)+\frac{1}{2}\right)\end{gather} up to second order in $g_{2}/v$. Within perturbation theory, the generating function is obtained as
\begin{widetext}
\begin{equation}
\label{eq:lng}\ln G(\lambda)=\frac{i\lambda}{\beta-i\lambda}\ln Z_{0}(\beta)+i\lambda E_{\mathrm{ad}}\left[1+\left(\frac{2\tau_{0}}{\beta-i\lambda}\right)^{2}\zeta\left(2,1+\frac{2\tau_{0}}{\beta-i\lambda}\right)\right]+E_{\mathrm{ad}}\tau_{0} h\left(\frac{\beta}{\tau_{0}},\frac{\lambda}{\tau_{0}}\right)
\end{equation}
\begin{equation}
\label{eq:hf}h\left(\frac{\beta}{\tau_{0}},\frac{\lambda}{\tau_{0}}\right)=1-\frac{1}{(z^{*})^{2}}\left[(1-z)\psi\left(\frac{1-z}{z^{*}} \right)+(1+z)\psi\left(\frac{1+z}{z^{*}}\right)-2\psi\left(\frac{1}{z^{*}} \right)\right]\qquad z=\frac{\beta+i\lambda}{2\tau_{0}}\,.
\end{equation}
\end{widetext}

In Eqs. \eqref{eq:lng} and \eqref{eq:hf}, $\zeta(x)$ is the generalized zeta function and $\psi(x)$ is the digamma function.
The partition function of the initial state is  $\ln Z_{0}(\beta)=L\pi/(6\beta v)$.
We note that $\ln G(\lambda)$ has poles on the lower complex semiplane only. It follows that the PDF is identically zero for energies lower than the ground state energy 
$E_{\mathrm{ad}}$, which meets physical expectations as well. Before presenting results on the PDF of total energy, we investigate two simple cases when the PDF can be calculated analytically and will also play important role later.

\jav{At zero temperature, Eq. \eqref{eq:lng} simplifies to
\begin{gather}
\ln G(\lambda;\beta\rightarrow\infty)=i\lambda E_{ad}-\frac{\lambda}{i\tau_0+\lambda}\alpha
\label{eq:gzt}
\end{gather}
which reproduces the results of Ref. \onlinecite{ztwork}, leading to a noncentral chi-squared distribution for the PDF.}

\jav{With finite initial temperature,} the behaviour of the unquenched case ($g_{2}=0$) is also interesting. The final state is the same as the initial thermal equilibrium state and the generating function reads as 
\begin{equation}
\label{eq:lngunpert}
\ln G(\lambda;g_{2}=0)=-\frac{\lambda}{i\beta+\lambda}\ln Z_{0}
\end{equation}
which leads to another noncentral chi-squared distribution with the noncentrality parameter $2\ln Z_{0}$. The PDF is
\begin{gather}
P(E;g_{2}=0)\label{eq:pdffintemp}
=\frac{e^{-\beta E}}{Z_{0}}\left[\delta(E)+\sqrt{\frac{L\pi}{6E v}}
I_{1}\left(2\sqrt{\frac{L\pi}{6 v} E}\right) \right],
\end{gather}
which is equal to the Boltzmann factor $\exp{(-\beta E)}/Z_{0}$ multiplied by the total energy density of states of a one-dimensional Bose gas with linear dispersion. 
The modified Bessel function behaves as an exponential function in the thermodynamic system limit and
almost all the spectral weight is carried by a \jav{non-Gaussian sharp peak centered at $\langle E\rangle=(\ln Z_{0})/\beta$ and of width $\Delta E=\sqrt{2\ln Z_{0}}/\beta$
as
\begin{gather}
P(E;g_2=0)\approx\frac{\beta(\ln Z_{0})^{\frac{1}{4}}}{2\sqrt{\pi}(\beta E)^{\frac{3}{4}}}\exp\left(-\left(\sqrt{\beta E}-\sqrt{\ln Z_{0}} \right)^2\right)
\end{gather}
whose high energy tail decays according to the Gamma distribution as $\exp(-\beta E)/(\beta E)^{3/4}$.}
\jav{In the strict thermodynamic limit, $L\rightarrow \infty$, the peak becomes infinitely narrow since $\Delta E/\langle E\rangle\rightarrow 0$, as universally expected, and}
the Dirac delta part of Eq. \eqref{eq:pdffintemp} is exponentially suppressed. For small systems, however, the Dirac-delta carries most of the probability weight and the continuous part contributes an \jav{exponentially decaying tail only
as
\begin{gather}
P(E;g_{2}=0)\approx\frac{1}{Z_{0}}\delta(E)+\frac{\beta\ln Z_{0}}{Z_{0}}\exp(-\beta E) \,.
\end{gather}}

Now we go on calculating the PDF of total energy within perturbation theory and at finite temperature.
At low temperature $\beta\gg\tau_{0}$, we obtain
\begin{equation}
\label{eq:sqlowT}
\ln G(\lambda;\beta \gg \tau_{0})=-\frac{\lambda}{i\beta+\lambda}\ln Z_{0}+i\lambda E_{\mathrm{ad}}-\frac{\lambda}{i\tau_{0}+\lambda}\alpha
\end{equation}
which is the sum of Eqs. \eqref{eq:gzt} and \eqref{eq:lngunpert}. This means that the PDF is the convolution of Eq. \eqref{eq:pdffintemp} and the zero temperature 
PDF both described by a noncentral chi-squared distribution. These consist of a Dirac delta part and a continuous part and so does their convolution. The weight of the Dirac delta after convolution is $\exp(-\alpha)/Z_{0}$ \jav{which is the probability of the vacuum-to-vacuum process}.

\begin{figure*}[t!]
\centering
\psfrag{PEpertau0}{$P(E)/\tau_{0}$}
\psfrag{Etau0}{$E\tau_{0}$}
\psfrag{alsm}{$\alpha=0.2$}
\psfrag{alim}{$\alpha=4$}
\psfrag{alla}{$\alpha=20$}
\pnt{-}
\pnt{0.0}\pnt{0.5}\pnt{1.0}\pnt{1.5}\pnt{2.0}\pnt{2.5}\pnt{0}\pnt{10}\pnt{20}\pnt{30}\pnt{40}\pnt{50}
\pnt{0.00}\pnt{0.01}\pnt{0.02}\pnt{0.03}\pnt{0.04}\pnt{0.05}\pnt{0.06}\pnt{0.07}\pnt{0.10}\pnt{0.15}\pnt{100}\pnt{150}\pnt{200}
a)\quad\includegraphics[scale=0.5]{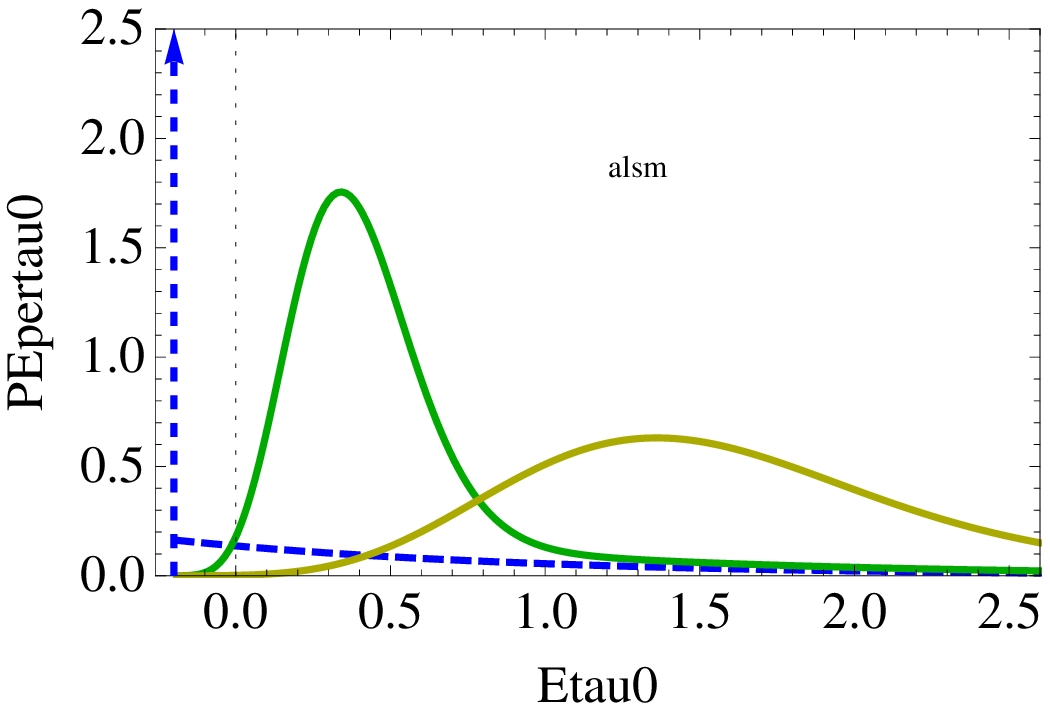}
b)\quad\includegraphics[scale=0.5]{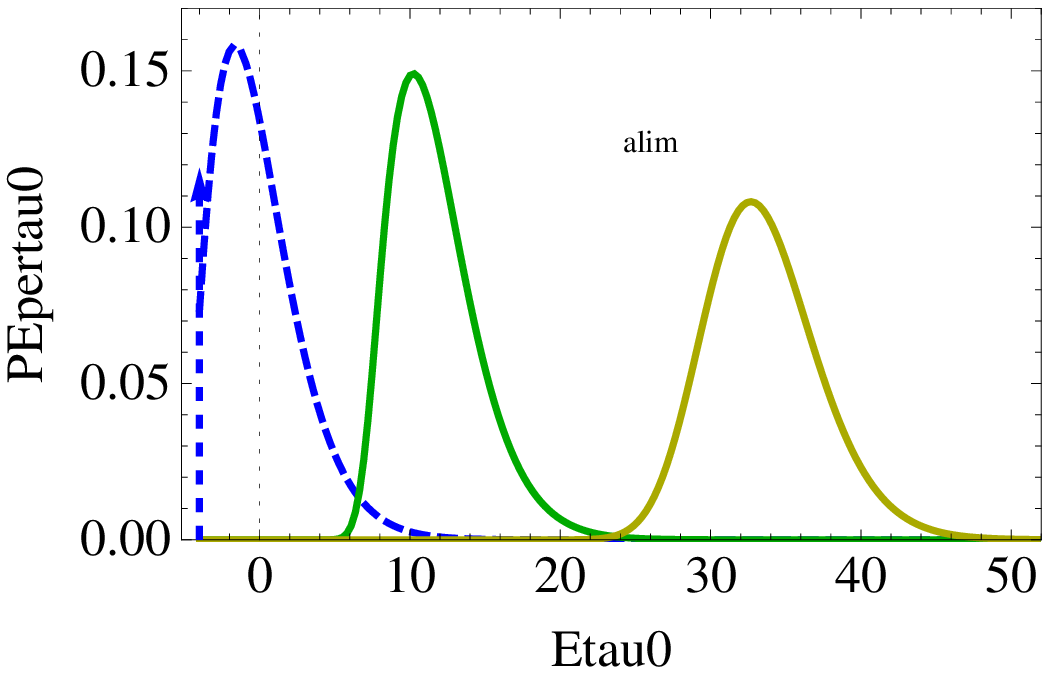}
c)\quad\includegraphics[scale=0.5]{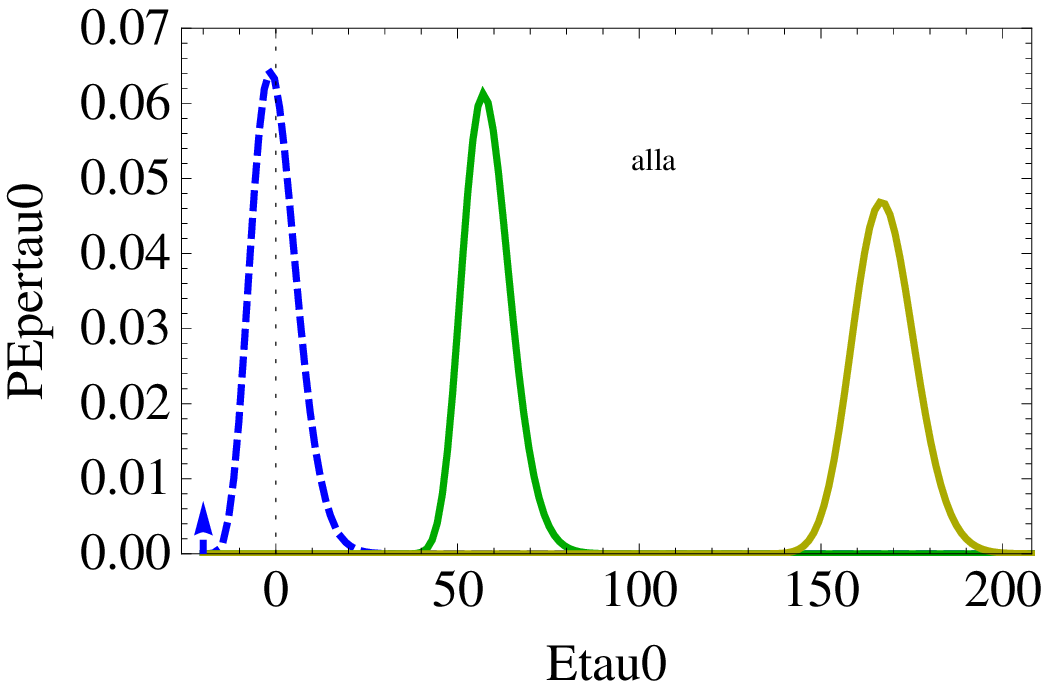}
\caption{Probability distribution function of total energy after SQ. In each figures the blue dashed curve corresponds to zero temperature. The analytic result\cite{ztwork} consists of a Dirac delta (blue dashed arrow) and a continuous part. Green and yellow curves show the continuous part of 
the total energy PDF at temperatures $\beta=30\tau_{0}$ and $\beta=10\tau_{0}$. The orthogonality exponent varies as: a) $\alpha=0.2$ small system b) $\alpha=4$ c) $\alpha=20$ large system. For all system sizes the distribution is shifted in positive direction and broadened as the temperature increases. For all figures $g_{2}/v=0.1$ was chosen.}
\label{fig:numpdfs}
\end{figure*}

We have numerically checked that at low temperature the convolution of the two abovementioned PDFs equals the exact PDF calculated by Fourier transforming the generating function Eq. \eqref{eq:lng}. 
The results of the numerical convolution are plotted in Fig. \ref{fig:numpdfs} for different system sizes. For small systems ($\alpha$ is small, see Fig. \ref{fig:numpdfs}.a), most of the spectral 
weight is carried by the Dirac-delta at zero temperature. With increasing temperature, one part of the Dirac delta stays at $E_{\mathrm{ad}}$ while another part evolves to a finite-width peak with the expectation value of $(\ln Z_{0})/\beta$.

In the thermodynamic limit (large system with $\alpha\gg 1$, see Fig. \ref{fig:numpdfs}.c) the zero temperature PDF consists of a \jav{ broadened peak at about $E=0$ and a Dirac delta at the adiabatic ground state energy
difference} with a small probability weight. 
At finite temperature the broadened  peak is shifted to $\langle E\rangle(T)=(\ln Z_{0})/\beta$ and its  width changes as $\Delta E(T)=\sqrt{2\alpha/\tau^{2}_{0}+2(\ln Z_{0})/\beta^{2}}$. 
At low temperatures \begin{gather}\frac{\Delta E(T)-\Delta E(0)}{\Delta E(0)}\sim T^{3}\,.\end{gather} This broadening is, however, not as spectacular as for small systems because its ratio 
with the shift of the peak scales as  $(\Delta E(T)-\Delta E(0))/\langle E\rangle(T)\approx 
\sqrt{2}\tau_{0}/(\sqrt{\alpha}\beta)$, being almost
negligible in the thermodynamic limit.

\section{Statistics of work}
\label{sec:wpdf}

Measuring the work statistics requires in principle two energy measurement\cite{hanggirmp,jarzynski}, one before and one after the time dependent protocol, though in the zero temperature case, the first one could be omitted\cite{ztwork}.
At finite temperature, however, this problem can be circumvented by coupling the system of interest to a qubit\cite{mazzola,dorner}, whose
interferometry or spectroscopy would yield the desired correlation function.
In the case of a LL, a hybrid system containing cold atoms and a flux qubit\cite{rmpnori} coupled to a Feshbach resonance was proposed at zero temperature to measure the Loschmidt echo using rf spectroscopy or Ramsey interferometry\cite{doraloschmidt}, and
this can readily be extended for finite temperatures as well. This setting  can also be useful to measure the statistics of work in other systems \cite{greekguy,yulia}.

In this section we investigate the PDF of work $\tilde{P}(W)$. At zero temperature this distribution coincides with the total energy distribution since the energy of the initial state is well-defined. 
At finite temperature, however, this is not the case. In the initial thermal equilibrium state the system can have arbitrary positive energy. It follows that the PDF of work differs from the PDF of total energy and has no lower bound.

The generating function of the distribution of work is defined as \cite{notanobservable} 
\begin{gather}
\tilde{G}(\lambda)=\mathrm{Tr}\left[\hat{\rho}_{0}e^{-i\lambda\hat{H}_{0}}e^{i\lambda\hat{H}_{H}(\tau)}\right],
\label{genfuncwork}
\end{gather}
 where $\hat{H}_{H}(\tau)$ is the final Hamiltonian in Heisenberg picture. We note that with finite 
temperature initial state the work statistics cannot be derived from the time dependent fidelity unlike the zero temperature case where the fidelity as a function of time and the generating function of work are basically the same in the case of a SQ\cite{silvaqcp}.

The generating function can still be rewritten by means of the generating function of the total energy $G(\lambda,\beta)$ as
\begin{gather}
\tilde{G}(\lambda,\beta)=\frac{Z_{0}(\beta+i\lambda)}{Z_{0}(\beta)}G(\lambda,\beta+i\lambda).
\end{gather}
Using Eq. \eqref{eq:lng}, we obtain
\begin{gather}
\ln \tilde{G}(\lambda)=i\lambda E_{\mathrm{ad}}\left(1+8\frac{\tau_{0}^{2}}{\beta^{2}}\zeta\left(2,1+\frac{2\tau_{0}}{\beta}\right) \right)+\nonumber \\
\label{eq:work}
+E_{\mathrm{ad}}\tau_{0} h\left(\frac{\beta+i\lambda}{\tau_{0}},\frac{\lambda}{\tau_{0}}\right)
\end{gather}
up to second order in $g_{2}/v$ where the function $h$ was defined in Eq. \eqref{eq:hf}. This result is valid only within perturbation theory and in the SQ limit. We note that $\ln\tilde{G}(\lambda)$ has poles on both complex semiplanes. This is the mathematical reason 
for the absence of lower bound in the PDF. It can be proven that Eq. \eqref{eq:work} satisfies the Jarzynski equality \cite{jarzynski}, i.e. $\langle \exp{(-\beta W)}\rangle= \tilde{G}(\lambda=i\beta)= Z(\tau)/Z_{0}$ where $Z(\tau)=\mathrm{Tr}\left[\exp{(-\beta \hat{H}(\tau))}\right]$ is 
the partition function of the final Hamiltonian.

\begin{figure*}[t!]
\psfrag{PWpertau0}{$P(W)/\tau_{0}$}
\psfrag{Wtau0}{$W\tau_{0}$}
\psfrag{alsm}{$\alpha=0.2$}
\psfrag{alim}{$\alpha=4$}
\psfrag{alla}{$\alpha=20$}
\pnt{-}
\pnt{20}\pnt{10}\pnt{0.0}\pnt{0.2}\pnt{0.4}\pnt{0.5}\pnt{1.0}\pnt{1.5}\pnt{2.0}\pnt{2.5}\pnt{3.0}\pnt{3.5}
\pnt{2}\pnt{4}\pnt{6}\pnt{8}\pnt{0}\pnt{0.00}\pnt{0.01}\pnt{0.02}\pnt{0.03}\pnt{0.04}\pnt{0.05}\pnt{0.06}\pnt{0.07}\pnt{0.10}\pnt{0.15}
\centering
a)\,\includegraphics[scale=0.5]{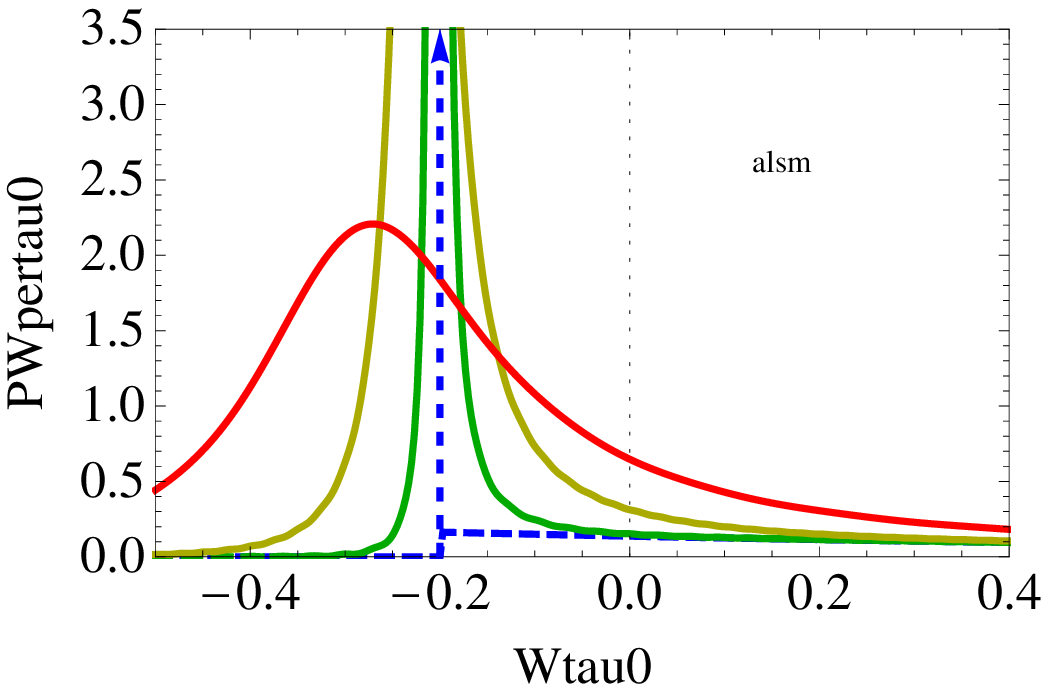}\quad
b)\,\includegraphics[scale=0.5]{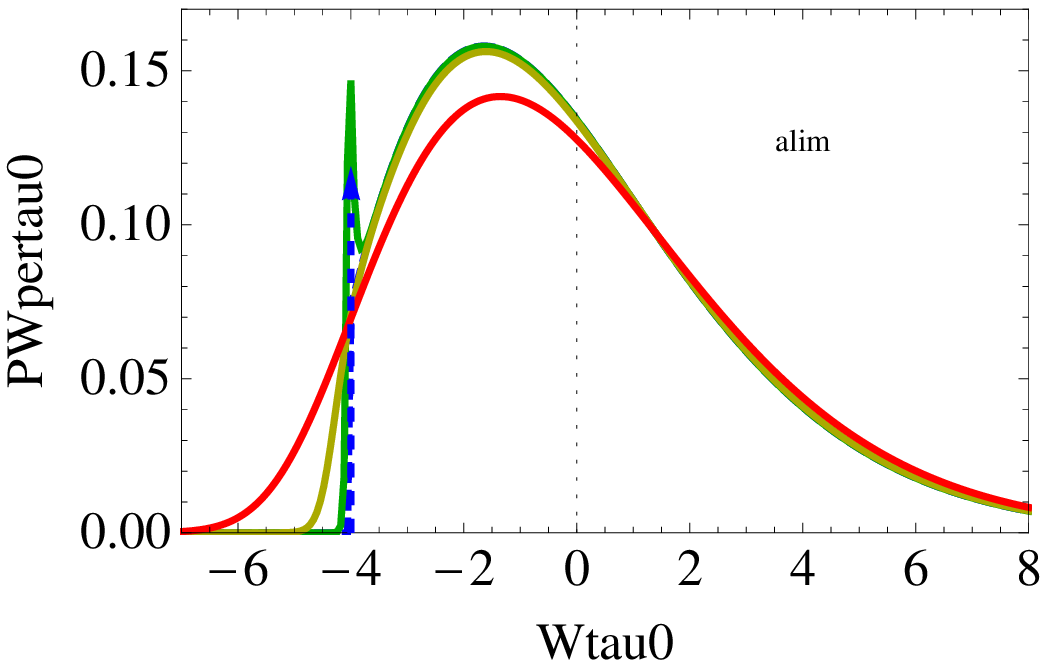}\quad
c)\,\includegraphics[scale=0.5]{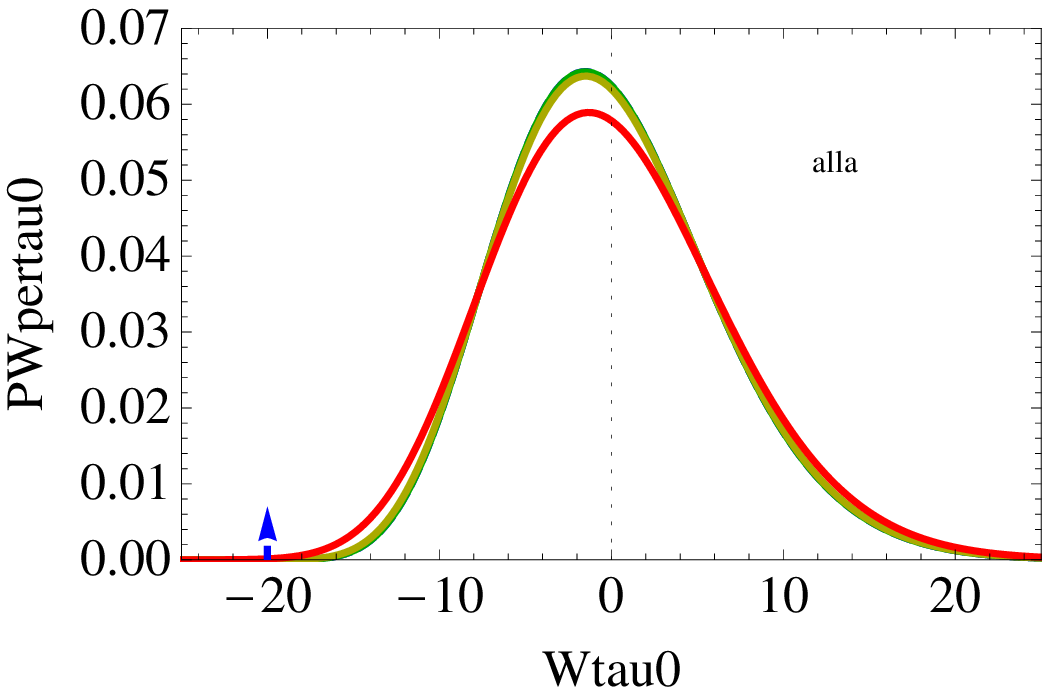}
\caption{Probability distribution function of work statistics after a SQ. In each figures the blue dashed curve shows the zero temperature case which was obtained analytically in our previous paper\cite{ztwork}. The blue arrow represent the Dirac delta part of the distributions. Green, yellow and red solid curves correspond to temperatures $\beta=30\tau_{0}$, $\beta=10\tau_{0}$ and $\beta=4\tau_{0}$, respectively. The orthogonality exponent varies as a) $\alpha=0.2$ small system, b) $\alpha=4$ c) $\alpha=20$ large system. a) In the small system limit most of the probability weight is carried by the Dirac delta part at zero temperature. At finite temperature this broadens to a finite width peak which dominates the distribution. c) In the large system limit the PDF of consists of a single peak which slightly broadens as the temperature increases. b) In the intermediate region one can notice the broadening of both the Dirac delta part and the continuous part of the PDF. 
The features of small and large system are combined in this region.}
\label{fig:workstat}
\end{figure*}

The Fourier transform of the generating function is evaluated numerically, and is shown in Fig. \ref{fig:workstat} for different system sizes and initial temperatures. 

The expectation value and the variance of the PDF of work is calculated analytically by taking the derivatives of the generating function Eq. \eqref{eq:work} at $\lambda=0$, yielding 
 $\langle W\rangle_{SQ}=0$ for arbitrary temperature and only the form  of the distribution changes as the temperature increases. 

In the small system limit,  the zero temperature PDF consists of a Dirac delta with large probability weight and an exponentially decaying tail (see Fig. \ref{fig:workstat}.a). At finite temperature the Dirac delta disappears and
 deforms to a broadened, negatively skewed peak. 
This deformation is a significant 
modification 
of the PDF.

In the thermodynamic  limit the only change is that the zero temperature peak slightly broadens (see Fig. \ref{fig:workstat}.c).
The width of the peak varies as
\begin{equation}
\label{eq:workwidth}\frac{\Delta W(T)-\Delta W(0)}{\Delta W(0)}=\frac{4\pi^{2}}{3}(\tau_{0}T)^{3},
\end{equation}
which means that significant broadening could be noticeable at higher temperatures only.

\section{Conclusion}
We have theoretically studied quantum quenches \jav{in the Luttinger model with finite temperature equilibrium initial state}. The steady state has been described by the diagonal ensemble, i.e. by determining the diagonal elements of the time evolved density matrix. At finite temperature, the boson numbers in the $+q$ and $-q$ modes can differ from each other and the difference may be larger for higher initial temperature.

The long time limit of the Uhlmann fidelity shows that the time evolved state deviates from the initial state with increasing temperature. 
How this relation depends on the quench duration is an interesting and open question.

Finite temperature effects in the statistics of final total energy and work done on the system during the quench have also been investigated in the SQ limit. It is worth mentioning again that these two distributions differ from each other, unlike in the zero temperature case, because in the 
finite 
temperature initial state the energy is not well defined. Within perturbation theory and at low temperature the PDF of total energy is found to be the convolution of the PDF of initial energy and the PDF of zero temperature work statistics. 
Numerical results show that the distribution is shifted and broadened due to finite temperature for both the thermodynamic and small system limits. In small systems, however, the broadening is more robust. 

The finite temperature effects in the statistics of work depend remarkably on the system size. In the small system limit,  significant rearrangement may be observed with increasing temperature, while in the thermodynamic limit, the peak of the PDF slightly broadens only. 
We believe that our results related to the distribution function of total energy and work done can be observed experimentally, using the setups suggested in Ref. \cite{dorner,mazzola,doraloschmidt}.

\begin{acknowledgments}
We thank M. Rigol and G. Zar\'and for stimulating comments. 
This work was supported by the Hungarian Scientific Research Fund under Grants No. OTKA K101244, K105149, CNK80991 and by the ERC Grant No. 259374-Sylo and by the Bolyai Program of the Hungarian Academy of Sciences.
\end{acknowledgments}

\bibliographystyle{apsrev}
\bibliography{fintempstatworklutt4}

\begin{thebibliography}{51}
\expandafter\ifx\csname natexlab\endcsname\relax\def\natexlab#1{#1}\fi
\expandafter\ifx\csname bibnamefont\endcsname\relax
  \def\bibnamefont#1{#1}\fi
\expandafter\ifx\csname bibfnamefont\endcsname\relax
  \def\bibfnamefont#1{#1}\fi
\expandafter\ifx\csname citenamefont\endcsname\relax
  \def\citenamefont#1{#1}\fi
\expandafter\ifx\csname url\endcsname\relax
  \def\url#1{\texttt{#1}}\fi
\expandafter\ifx\csname urlprefix\endcsname\relax\def\urlprefix{URL }\fi
\providecommand{\bibinfo}[2]{#2}
\providecommand{\eprint}[2][]{\url{#2}}

\bibitem[{\citenamefont{Greiner et~al.}(2002)\citenamefont{Greiner, Mandel,
  H{\"a}nsch, and Bloch}}]{greinernat}
\bibinfo{author}{\bibfnamefont{M.}~\bibnamefont{Greiner}},
  \bibinfo{author}{\bibfnamefont{O.}~\bibnamefont{Mandel}},
  \bibinfo{author}{\bibfnamefont{T.~W.} \bibnamefont{H{\"a}nsch}},
  \bibnamefont{and} \bibinfo{author}{\bibfnamefont{I.}~\bibnamefont{Bloch}},
  \bibinfo{journal}{Nature} \textbf{\bibinfo{volume}{419}}, \bibinfo{pages}{51}
  (\bibinfo{year}{2002}).

\bibitem[{\citenamefont{Kinoshita et~al.}(2006)\citenamefont{Kinoshita, Wenger,
  and Weiss}}]{toshiyanat}
\bibinfo{author}{\bibfnamefont{T.}~\bibnamefont{Kinoshita}},
  \bibinfo{author}{\bibfnamefont{T.}~\bibnamefont{Wenger}}, \bibnamefont{and}
  \bibinfo{author}{\bibfnamefont{D.~S.} \bibnamefont{Weiss}},
  \bibinfo{journal}{Nature} \textbf{\bibinfo{volume}{440}},
  \bibinfo{pages}{900} (\bibinfo{year}{2006}).

\bibitem[{\citenamefont{Hofferberth et~al.}(2007)\citenamefont{Hofferberth,
  Lesanovsky, Fischer, Schumm, and Schmiedmayer}}]{hofferberthnat}
\bibinfo{author}{\bibfnamefont{S.}~\bibnamefont{Hofferberth}},
  \bibinfo{author}{\bibfnamefont{I.}~\bibnamefont{Lesanovsky}},
  \bibinfo{author}{\bibfnamefont{B.}~\bibnamefont{Fischer}},
  \bibinfo{author}{\bibfnamefont{T.}~\bibnamefont{Schumm}}, \bibnamefont{and}
  \bibinfo{author}{\bibfnamefont{J.}~\bibnamefont{Schmiedmayer}},
  \bibinfo{journal}{Nature} \textbf{\bibinfo{volume}{449}},
  \bibinfo{pages}{324} (\bibinfo{year}{2007}).

\bibitem[{\citenamefont{Bloch et~al.}(2008)\citenamefont{Bloch, Dalibard, and
  Zwerger}}]{blochrmp}
\bibinfo{author}{\bibfnamefont{I.}~\bibnamefont{Bloch}},
  \bibinfo{author}{\bibfnamefont{J.}~\bibnamefont{Dalibard}}, \bibnamefont{and}
  \bibinfo{author}{\bibfnamefont{W.}~\bibnamefont{Zwerger}},
  \bibinfo{journal}{Rev. Mod. Phys.} \textbf{\bibinfo{volume}{80}},
  \bibinfo{pages}{885} (\bibinfo{year}{2008}).

\bibitem[{\citenamefont{Dziarmaga}(2010)}]{dziarmagareview}
\bibinfo{author}{\bibfnamefont{J.}~\bibnamefont{Dziarmaga}},
  \bibinfo{journal}{Adv. Phys.} \textbf{\bibinfo{volume}{59}},
  \bibinfo{pages}{1063} (\bibinfo{year}{2010}).

\bibitem[{\citenamefont{Rigol}(2009)}]{rigoltherm2}
\bibinfo{author}{\bibfnamefont{M.}~\bibnamefont{Rigol}},
  \bibinfo{journal}{Phys. Rev. Lett.} \textbf{\bibinfo{volume}{103}},
  \bibinfo{pages}{100403} (\bibinfo{year}{2009}).

\bibitem[{\citenamefont{Polkovnikov et~al.}(2011)\citenamefont{Polkovnikov,
  Sengupta, Silva, and Vengalattore}}]{polkovnikovrmp}
\bibinfo{author}{\bibfnamefont{A.}~\bibnamefont{Polkovnikov}},
  \bibinfo{author}{\bibfnamefont{K.}~\bibnamefont{Sengupta}},
  \bibinfo{author}{\bibfnamefont{A.}~\bibnamefont{Silva}}, \bibnamefont{and}
  \bibinfo{author}{\bibfnamefont{M.}~\bibnamefont{Vengalattore}},
  \bibinfo{journal}{Rev. Mod. Phys.} \textbf{\bibinfo{volume}{83}},
  \bibinfo{pages}{863} (\bibinfo{year}{2011}).

\bibitem[{\citenamefont{Cassidy et~al.}(2011)\citenamefont{Cassidy, Clark, and
  Rigol}}]{rigoltherm}
\bibinfo{author}{\bibfnamefont{A.~C.} \bibnamefont{Cassidy}},
  \bibinfo{author}{\bibfnamefont{C.~W.} \bibnamefont{Clark}}, \bibnamefont{and}
  \bibinfo{author}{\bibfnamefont{M.}~\bibnamefont{Rigol}},
  \bibinfo{journal}{Phys. Rev. Lett.} \textbf{\bibinfo{volume}{106}},
  \bibinfo{pages}{140405} (\bibinfo{year}{2011}).

\bibitem[{\citenamefont{Rigol and Srednicki}(2012)}]{rigoleth}
\bibinfo{author}{\bibfnamefont{M.}~\bibnamefont{Rigol}} \bibnamefont{and}
  \bibinfo{author}{\bibfnamefont{M.}~\bibnamefont{Srednicki}},
  \bibinfo{journal}{Phys. Rev. Lett.} \textbf{\bibinfo{volume}{108}},
  \bibinfo{pages}{110601} (\bibinfo{year}{2012}).

\bibitem[{\citenamefont{Rigol et~al.}(2007)\citenamefont{Rigol, Dunjko,
  Yurovsky, and Olshanii}}]{rigolgge}
\bibinfo{author}{\bibfnamefont{M.}~\bibnamefont{Rigol}},
  \bibinfo{author}{\bibfnamefont{V.}~\bibnamefont{Dunjko}},
  \bibinfo{author}{\bibfnamefont{V.}~\bibnamefont{Yurovsky}}, \bibnamefont{and}
  \bibinfo{author}{\bibfnamefont{M.}~\bibnamefont{Olshanii}},
  \bibinfo{journal}{Phys. Rev. Lett.} \textbf{\bibinfo{volume}{98}},
  \bibinfo{pages}{050405} (\bibinfo{year}{2007}).

\bibitem[{\citenamefont{Rigol et~al.}(2008)\citenamefont{Rigol, Dunjko, and
  Olshanii}}]{rigolnat}
\bibinfo{author}{\bibfnamefont{M.}~\bibnamefont{Rigol}},
  \bibinfo{author}{\bibfnamefont{V.}~\bibnamefont{Dunjko}}, \bibnamefont{and}
  \bibinfo{author}{\bibfnamefont{M.}~\bibnamefont{Olshanii}},
  \bibinfo{journal}{Nature} \textbf{\bibinfo{volume}{452}},
  \bibinfo{pages}{854} (\bibinfo{year}{2008}).

\bibitem[{\citenamefont{Caux and Konik}(2012)}]{ggecaux}
\bibinfo{author}{\bibfnamefont{J.-S.} \bibnamefont{Caux}} \bibnamefont{and}
  \bibinfo{author}{\bibfnamefont{R.~M.} \bibnamefont{Konik}},
  \bibinfo{journal}{Phys. Rev. Lett.} \textbf{\bibinfo{volume}{109}},
  \bibinfo{pages}{175301} (\bibinfo{year}{2012}).

\bibitem[{\citenamefont{Cazalilla et~al.}(2011)\citenamefont{Cazalilla, Citro,
  Giamarchi, Orignac, and Rigol}}]{cazalillarmp}
\bibinfo{author}{\bibfnamefont{M.~A.} \bibnamefont{Cazalilla}},
  \bibinfo{author}{\bibfnamefont{R.}~\bibnamefont{Citro}},
  \bibinfo{author}{\bibfnamefont{T.}~\bibnamefont{Giamarchi}},
  \bibinfo{author}{\bibfnamefont{E.}~\bibnamefont{Orignac}}, \bibnamefont{and}
  \bibinfo{author}{\bibfnamefont{M.}~\bibnamefont{Rigol}},
  \bibinfo{journal}{Rev. Mod. Phys.} \textbf{\bibinfo{volume}{83}},
  \bibinfo{pages}{1405} (\bibinfo{year}{2011}).

\bibitem[{\citenamefont{Karrasch et~al.}(2012)\citenamefont{Karrasch, Rentrop,
  Schuricht, and Meden}}]{medenprl}
\bibinfo{author}{\bibfnamefont{C.}~\bibnamefont{Karrasch}},
  \bibinfo{author}{\bibfnamefont{J.}~\bibnamefont{Rentrop}},
  \bibinfo{author}{\bibfnamefont{D.}~\bibnamefont{Schuricht}},
  \bibnamefont{and} \bibinfo{author}{\bibfnamefont{V.}~\bibnamefont{Meden}},
  \bibinfo{journal}{Phys. Rev. Lett.} \textbf{\bibinfo{volume}{109}},
  \bibinfo{pages}{126406} (\bibinfo{year}{2012}).

\bibitem[{\citenamefont{Pollmann et~al.}(2013)\citenamefont{Pollmann, Haque,
  and D\'ora}}]{balazslattnum}
\bibinfo{author}{\bibfnamefont{F.}~\bibnamefont{Pollmann}},
  \bibinfo{author}{\bibfnamefont{M.}~\bibnamefont{Haque}}, \bibnamefont{and}
  \bibinfo{author}{\bibfnamefont{B.}~\bibnamefont{D\'ora}},
  \bibinfo{journal}{Phys. Rev. B} \textbf{\bibinfo{volume}{87}},
  \bibinfo{pages}{041109} (\bibinfo{year}{2013}).

\bibitem[{\citenamefont{D\'ora et~al.}(2013)\citenamefont{D\'ora, Pollmann,
  Fort\'agh, and Zar\'and}}]{doraloschmidt}
\bibinfo{author}{\bibfnamefont{B.}~\bibnamefont{D\'ora}},
  \bibinfo{author}{\bibfnamefont{F.}~\bibnamefont{Pollmann}},
  \bibinfo{author}{\bibfnamefont{J.}~\bibnamefont{Fort\'agh}},
  \bibnamefont{and} \bibinfo{author}{\bibfnamefont{G.}~\bibnamefont{Zar\'and}},
  \bibinfo{journal}{Phys. Rev. Lett.} \textbf{\bibinfo{volume}{111}},
  \bibinfo{pages}{046402} (\bibinfo{year}{2013}).

\bibitem[{\citenamefont{Cazalilla}(2006)}]{cazalillaprl}
\bibinfo{author}{\bibfnamefont{M.~A.} \bibnamefont{Cazalilla}},
  \bibinfo{journal}{Phys. Rev. Lett.} \textbf{\bibinfo{volume}{97}},
  \bibinfo{pages}{156403} (\bibinfo{year}{2006}).

\bibitem[{\citenamefont{Iucci and Cazalilla}(2009)}]{cazalillapra}
\bibinfo{author}{\bibfnamefont{A.}~\bibnamefont{Iucci}} \bibnamefont{and}
  \bibinfo{author}{\bibfnamefont{M.~A.} \bibnamefont{Cazalilla}},
  \bibinfo{journal}{Phys. Rev. A} \textbf{\bibinfo{volume}{80}},
  \bibinfo{pages}{063619} (\bibinfo{year}{2009}).

\bibitem[{\citenamefont{D\'ora et~al.}(2011)\citenamefont{D\'ora, Haque, and
  Zar\'and}}]{balazsprl}
\bibinfo{author}{\bibfnamefont{B.}~\bibnamefont{D\'ora}},
  \bibinfo{author}{\bibfnamefont{M.}~\bibnamefont{Haque}}, \bibnamefont{and}
  \bibinfo{author}{\bibfnamefont{G.}~\bibnamefont{Zar\'and}},
  \bibinfo{journal}{Phys. Rev. Lett.} \textbf{\bibinfo{volume}{106}},
  \bibinfo{pages}{156406} (\bibinfo{year}{2011}).

\bibitem[{\citenamefont{Perfetto and Stefanucci}(2011)}]{perfetto}
\bibinfo{author}{\bibfnamefont{E.}~\bibnamefont{Perfetto}} \bibnamefont{and}
  \bibinfo{author}{\bibfnamefont{G.}~\bibnamefont{Stefanucci}},
  \bibinfo{journal}{Eur. Phys. Lett.} \textbf{\bibinfo{volume}{95}},
  \bibinfo{pages}{10006} (\bibinfo{year}{2011}).

\bibitem[{\citenamefont{Mitra and Giamarchi}(2011)}]{mitra}
\bibinfo{author}{\bibfnamefont{A.}~\bibnamefont{Mitra}} \bibnamefont{and}
  \bibinfo{author}{\bibfnamefont{T.}~\bibnamefont{Giamarchi}},
  \bibinfo{journal}{Phys. Rev. Lett.} \textbf{\bibinfo{volume}{107}},
  \bibinfo{pages}{150602} (\bibinfo{year}{2011}).

\bibitem[{\citenamefont{Mitra}(2012)}]{mitra1}
\bibinfo{author}{\bibfnamefont{A.}~\bibnamefont{Mitra}},
  \bibinfo{journal}{Phys. Rev. Lett.} \textbf{\bibinfo{volume}{109}},
  \bibinfo{pages}{260601} (\bibinfo{year}{2012}).

\bibitem[{\citenamefont{Nessi and Iucci}(2013)}]{nessiprb}
\bibinfo{author}{\bibfnamefont{N.}~\bibnamefont{Nessi}} \bibnamefont{and}
  \bibinfo{author}{\bibfnamefont{A.}~\bibnamefont{Iucci}},
  \bibinfo{journal}{Phys. Rev. B} \textbf{\bibinfo{volume}{87}},
  \bibinfo{pages}{085137} (\bibinfo{year}{2013}).

\bibitem[{\citenamefont{D\'ora et~al.}(2012)\citenamefont{D\'ora, B\'acsi, and
  Zar\'and}}]{ztwork}
\bibinfo{author}{\bibfnamefont{B.}~\bibnamefont{D\'ora}},
  \bibinfo{author}{\bibfnamefont{{\'A}.}~\bibnamefont{B\'acsi}},
  \bibnamefont{and} \bibinfo{author}{\bibfnamefont{G.}~\bibnamefont{Zar\'and}},
  \bibinfo{journal}{Phys. Rev. B} \textbf{\bibinfo{volume}{86}},
  \bibinfo{pages}{161109} (\bibinfo{year}{2012}).

\bibitem[{\citenamefont{He and Rigol}(2012)}]{rigolfintemp}
\bibinfo{author}{\bibfnamefont{K.}~\bibnamefont{He}} \bibnamefont{and}
  \bibinfo{author}{\bibfnamefont{M.}~\bibnamefont{Rigol}},
  \bibinfo{journal}{Phys. Rev. A} \textbf{\bibinfo{volume}{85}},
  \bibinfo{pages}{063609} (\bibinfo{year}{2012}).

\bibitem[{\citenamefont{Dziarmaga and Tylutki}(2011)}]{tylutki}
\bibinfo{author}{\bibfnamefont{J.}~\bibnamefont{Dziarmaga}} \bibnamefont{and}
  \bibinfo{author}{\bibfnamefont{M.}~\bibnamefont{Tylutki}},
  \bibinfo{journal}{Phys. Rev. B} \textbf{\bibinfo{volume}{84}},
  \bibinfo{pages}{214522} (\bibinfo{year}{2011}).

\bibitem[{\citenamefont{Venuti and Zanardi}(2010)}]{zanardipra}
\bibinfo{author}{\bibfnamefont{L.~C.} \bibnamefont{Venuti}} \bibnamefont{and}
  \bibinfo{author}{\bibfnamefont{P.}~\bibnamefont{Zanardi}},
  \bibinfo{journal}{Phys. Rev. A} \textbf{\bibinfo{volume}{81}},
  \bibinfo{pages}{022113} (\bibinfo{year}{2010}).

\bibitem[{\citenamefont{Jacobson et~al.}(2011)\citenamefont{Jacobson, Venuti,
  and Zanardi}}]{zanardi}
\bibinfo{author}{\bibfnamefont{N.~T.} \bibnamefont{Jacobson}},
  \bibinfo{author}{\bibfnamefont{L.~C.} \bibnamefont{Venuti}},
  \bibnamefont{and} \bibinfo{author}{\bibfnamefont{P.}~\bibnamefont{Zanardi}},
  \bibinfo{journal}{Phys. Rev. A} \textbf{\bibinfo{volume}{84}},
  \bibinfo{pages}{022115} (\bibinfo{year}{2011}).

\bibitem[{\citenamefont{Rams and Damski}(2011)}]{damskiprl}
\bibinfo{author}{\bibfnamefont{M.~M.} \bibnamefont{Rams}} \bibnamefont{and}
  \bibinfo{author}{\bibfnamefont{B.}~\bibnamefont{Damski}},
  \bibinfo{journal}{Phys. Rev. Lett.} \textbf{\bibinfo{volume}{106}},
  \bibinfo{pages}{055701} (\bibinfo{year}{2011}).

\bibitem[{\citenamefont{Venuti et~al.}(2011)\citenamefont{Venuti, Jacobson,
  Santra, and Zanardi}}]{zanardifintemp}
\bibinfo{author}{\bibfnamefont{L.~C.} \bibnamefont{Venuti}},
  \bibinfo{author}{\bibfnamefont{N.~T.} \bibnamefont{Jacobson}},
  \bibinfo{author}{\bibfnamefont{S.}~\bibnamefont{Santra}}, \bibnamefont{and}
  \bibinfo{author}{\bibfnamefont{P.}~\bibnamefont{Zanardi}},
  \bibinfo{journal}{Phys. Rev. Lett.} \textbf{\bibinfo{volume}{107}},
  \bibinfo{pages}{010403} (\bibinfo{year}{2011}).

\bibitem[{\citenamefont{Gorin et~al.}(2006)\citenamefont{Gorin, Prosen,
  Seligman, and Znidaric}}]{Gorin200633}
\bibinfo{author}{\bibfnamefont{T.}~\bibnamefont{Gorin}},
  \bibinfo{author}{\bibfnamefont{T.}~\bibnamefont{Prosen}},
  \bibinfo{author}{\bibfnamefont{T.~H.} \bibnamefont{Seligman}},
  \bibnamefont{and} \bibinfo{author}{\bibfnamefont{M.}~\bibnamefont{Znidaric}},
  \bibinfo{journal}{Physics Reports} \textbf{\bibinfo{volume}{435}},
  \bibinfo{pages}{33 } (\bibinfo{year}{2006}).

\bibitem[{\citenamefont{Goussev et~al.}(2012)\citenamefont{Goussev, Jalabert,
  Pastawski, and Wisniacki}}]{goussevscholar}
\bibinfo{author}{\bibfnamefont{A.}~\bibnamefont{Goussev}},
  \bibinfo{author}{\bibfnamefont{R.~A.} \bibnamefont{Jalabert}},
  \bibinfo{author}{\bibfnamefont{H.~M.} \bibnamefont{Pastawski}},
  \bibnamefont{and} \bibinfo{author}{\bibfnamefont{D.~A.}
  \bibnamefont{Wisniacki}}, \bibinfo{journal}{Scholarpedia}
  \textbf{\bibinfo{volume}{7}}, \bibinfo{pages}{11687} (\bibinfo{year}{2012}).

\bibitem[{\citenamefont{Silva}(2008)}]{silvaqcp}
\bibinfo{author}{\bibfnamefont{A.}~\bibnamefont{Silva}},
  \bibinfo{journal}{Phys. Rev. Lett.} \textbf{\bibinfo{volume}{101}},
  \bibinfo{pages}{120603} (\bibinfo{year}{2008}).

\bibitem[{\citenamefont{Campisi et~al.}(2011)\citenamefont{Campisi, H{\"a}nggi,
  and Talkner}}]{hanggirmp}
\bibinfo{author}{\bibfnamefont{M.}~\bibnamefont{Campisi}},
  \bibinfo{author}{\bibfnamefont{P.}~\bibnamefont{H{\"a}nggi}},
  \bibnamefont{and} \bibinfo{author}{\bibfnamefont{P.}~\bibnamefont{Talkner}},
  \bibinfo{journal}{Rev. Mod. Phys.} \textbf{\bibinfo{volume}{83}},
  \bibinfo{pages}{771} (\bibinfo{year}{2011}).

\bibitem[{\citenamefont{Jarzynski}(1997)}]{jarzynski}
\bibinfo{author}{\bibfnamefont{C.}~\bibnamefont{Jarzynski}},
  \bibinfo{journal}{Phys. Rev. Lett.} \textbf{\bibinfo{volume}{78}},
  \bibinfo{pages}{2690} (\bibinfo{year}{1997}).

\bibitem[{\citenamefont{Giamarchi}(2004)}]{giamarchi}
\bibinfo{author}{\bibfnamefont{T.}~\bibnamefont{Giamarchi}},
  \emph{\bibinfo{title}{Quantum Physics in One Dimension}}
  (\bibinfo{publisher}{Oxford University Press}, \bibinfo{address}{Oxford},
  \bibinfo{year}{2004}).

\bibitem[{\citenamefont{Gogolin et~al.}(1998)\citenamefont{Gogolin, Nersesyan,
  and Tsvelik}}]{nersesyan}
\bibinfo{author}{\bibfnamefont{A.~O.} \bibnamefont{Gogolin}},
  \bibinfo{author}{\bibfnamefont{A.~A.} \bibnamefont{Nersesyan}},
  \bibnamefont{and} \bibinfo{author}{\bibfnamefont{A.~M.}
  \bibnamefont{Tsvelik}}, \emph{\bibinfo{title}{Bosonization and Strongly
  Correlated Systems}} (\bibinfo{publisher}{Cambridge University Press},
  \bibinfo{address}{Cambridge}, \bibinfo{year}{1998}).

\bibitem[{\citenamefont{Kennes and Meden}()}]{kennes}
\bibinfo{author}{\bibfnamefont{D.~M.} \bibnamefont{Kennes}} \bibnamefont{and}
  \bibinfo{author}{\bibfnamefont{V.}~\bibnamefont{Meden}},
  \bibinfo{note}{{a}rXiv:1304.5889}.

\bibitem[{\citenamefont{Barthel and Schollw\"ock}(2008)}]{barthel}
\bibinfo{author}{\bibfnamefont{T.}~\bibnamefont{Barthel}} \bibnamefont{and}
  \bibinfo{author}{\bibfnamefont{U.}~\bibnamefont{Schollw\"ock}},
  \bibinfo{journal}{Phys. Rev. Lett.} \textbf{\bibinfo{volume}{100}},
  \bibinfo{pages}{100601} (\bibinfo{year}{2008}).

\bibitem[{\citenamefont{Peres}(1984)}]{aperes}
\bibinfo{author}{\bibfnamefont{A.}~\bibnamefont{Peres}},
  \bibinfo{journal}{Phys. Rev. A} \textbf{\bibinfo{volume}{30}},
  \bibinfo{pages}{1610} (\bibinfo{year}{1984}).

\bibitem[{\citenamefont{Jozsa}(1994)}]{jozsauhlmann}
\bibinfo{author}{\bibfnamefont{R.}~\bibnamefont{Jozsa}}, \bibinfo{journal}{J.
  Mod. Opt.} \textbf{\bibinfo{volume}{41}}, \bibinfo{pages}{2315}
  (\bibinfo{year}{1994}).

\bibitem[{\citenamefont{Nielsen and Chuang}(2010)}]{nielsen}
\bibinfo{author}{\bibfnamefont{M.~A.} \bibnamefont{Nielsen}} \bibnamefont{and}
  \bibinfo{author}{\bibfnamefont{I.~L.} \bibnamefont{Chuang}},
  \emph{\bibinfo{title}{Quantum computation and quantum information}}
  (\bibinfo{publisher}{Cambridge University Press},
  \bibinfo{address}{Cambridge}, \bibinfo{year}{2010}).

\bibitem[{\citenamefont{Sirker}(2010)}]{sirker}
\bibinfo{author}{\bibfnamefont{J.}~\bibnamefont{Sirker}},
  \bibinfo{journal}{Phys. Rev. Lett.} \textbf{\bibinfo{volume}{105}},
  \bibinfo{pages}{117203} (\bibinfo{year}{2010}).

\bibitem[{\citenamefont{Mazzola et~al.}(2013)\citenamefont{Mazzola, De~Chiara,
  and Paternostro}}]{mazzola}
\bibinfo{author}{\bibfnamefont{L.}~\bibnamefont{Mazzola}},
  \bibinfo{author}{\bibfnamefont{G.}~\bibnamefont{De~Chiara}},
  \bibnamefont{and}
  \bibinfo{author}{\bibfnamefont{M.}~\bibnamefont{Paternostro}},
  \bibinfo{journal}{Phys. Rev. Lett.} \textbf{\bibinfo{volume}{110}},
  \bibinfo{pages}{230602} (\bibinfo{year}{2013}).

\bibitem[{\citenamefont{Dorner et~al.}(2013)\citenamefont{Dorner, Clark,
  Heaney, Fazio, Goold, and Vedral}}]{dorner}
\bibinfo{author}{\bibfnamefont{R.}~\bibnamefont{Dorner}},
  \bibinfo{author}{\bibfnamefont{S.~R.} \bibnamefont{Clark}},
  \bibinfo{author}{\bibfnamefont{L.}~\bibnamefont{Heaney}},
  \bibinfo{author}{\bibfnamefont{R.}~\bibnamefont{Fazio}},
  \bibinfo{author}{\bibfnamefont{J.}~\bibnamefont{Goold}}, \bibnamefont{and}
  \bibinfo{author}{\bibfnamefont{V.}~\bibnamefont{Vedral}},
  \bibinfo{journal}{Phys. Rev. Lett.} \textbf{\bibinfo{volume}{110}},
  \bibinfo{pages}{230601} (\bibinfo{year}{2013}).

\bibitem[{\citenamefont{Xiang et~al.}(2013)\citenamefont{Xiang, Ashhab, You,
  and Nori}}]{rmpnori}
\bibinfo{author}{\bibfnamefont{Z.-L.} \bibnamefont{Xiang}},
  \bibinfo{author}{\bibfnamefont{S.}~\bibnamefont{Ashhab}},
  \bibinfo{author}{\bibfnamefont{J.~Q.} \bibnamefont{You}}, \bibnamefont{and}
  \bibinfo{author}{\bibfnamefont{F.}~\bibnamefont{Nori}},
  \bibinfo{journal}{Rev. Mod. Phys.} \textbf{\bibinfo{volume}{85}},
  \bibinfo{pages}{623} (\bibinfo{year}{2013}).

\bibitem[{\citenamefont{Sotiriadis et~al.}(2013)\citenamefont{Sotiriadis,
  Gambassi, and Silva}}]{greekguy}
\bibinfo{author}{\bibfnamefont{S.}~\bibnamefont{Sotiriadis}},
  \bibinfo{author}{\bibfnamefont{A.}~\bibnamefont{Gambassi}}, \bibnamefont{and}
  \bibinfo{author}{\bibfnamefont{A.}~\bibnamefont{Silva}},
  \bibinfo{journal}{Phys. Rev. E} \textbf{\bibinfo{volume}{87}},
  \bibinfo{pages}{052129} (\bibinfo{year}{2013}).

\bibitem[{\citenamefont{Shchadilova et~al.}()\citenamefont{Shchadilova,
  Ribeiro, and Haque}}]{yulia}
\bibinfo{author}{\bibfnamefont{Y.~E.} \bibnamefont{Shchadilova}},
  \bibinfo{author}{\bibfnamefont{P.}~\bibnamefont{Ribeiro}}, \bibnamefont{and}
  \bibinfo{author}{\bibfnamefont{M.}~\bibnamefont{Haque}},
  \bibinfo{note}{{a}rXiv:1303.4103}.

\bibitem[{\citenamefont{Talkner et~al.}(2007)\citenamefont{Talkner, Lutz, and
  H{\"a}nggi}}]{notanobservable}
\bibinfo{author}{\bibfnamefont{P.}~\bibnamefont{Talkner}},
  \bibinfo{author}{\bibfnamefont{E.}~\bibnamefont{Lutz}}, \bibnamefont{and}
  \bibinfo{author}{\bibfnamefont{P.}~\bibnamefont{H{\"a}nggi}},
  \bibinfo{journal}{Phys. Rev. E} \textbf{\bibinfo{volume}{75}},
  \bibinfo{pages}{050102} (\bibinfo{year}{2007}).

\bibitem[{\citenamefont{Solomon}(1971)}]{solomon}
\bibinfo{author}{\bibfnamefont{A.~I.} \bibnamefont{Solomon}},
  \bibinfo{journal}{J. Math. Phys.} \textbf{\bibinfo{volume}{12}},
  \bibinfo{pages}{390} (\bibinfo{year}{1971}).

\bibitem[{\citenamefont{Gilmore}(1974)}]{gilmore}
\bibinfo{author}{\bibfnamefont{R.}~\bibnamefont{Gilmore}}, \bibinfo{journal}{J.
  Math. Phys.} \textbf{\bibinfo{volume}{15}}, \bibinfo{pages}{2090}
  (\bibinfo{year}{1974}).

\end{thebibliography}

\appendix
\section{Derivation of the generating function of occupation number distribution}
\label{sec:Fxi}
\noindent The generating function of the occupation number probability distribution is defined as
\begin{gather}
\label{eq:fdef2}
f(\xi_{+},\xi_{-})=\mathrm{Tr}\,\left[\hat{\rho}(\tau)e^{i(\xi_{+}\hat{n}_{+}+\xi_{-}\hat{n}_{-})}\right]
\end{gather}
where $\hat{\rho}(\tau)$ is the exact time evolved density operator after the quench, see Eq. \eqref{eq:fdef}. In this section we consider only a single $q>0$ mode. The initial state is 
\begin{gather}\hat{\rho}_{0}=\frac{e^{-\beta\omega_{0}(b_{+}^{+}b^{}_{+}+b_{-}^{+}b^{}_{-})}}{z_{0}}\qquad\qquad z_{0}=(1-e^{-\beta\omega_{0}})^{-2}\end{gather}
describing a canonical ensemble.
We introduce the operators
\begin{gather}\hat{K}_{0}=\frac{d_{+}^{+}d_{+}+d_{-}d_{-}^{+}}{2}\end{gather}
\begin{gather}\hat{K}_{+}=d^{+}_{+}d^{+}_{-}\qquad \hat{K}_{-}=d_{+}d_{-}=\hat{K}_{+}^{+}\end{gather}
where $d_{\pm}$ is the annihilation operator of quasiparticles diagonalizing the final Hamiltonian. The operator $\hat{K}_{0}$ does not change the number of bosons while $\hat{K}_{+}$ ($\hat{K}_{-}$) creates (annihilates) a pair of $d_{+}$ and $d_{-}$ bosons. The operators obey the commutation 
relations of $su(1,1)$ algebra, $\left[\hat{K}_{0},\hat{K}_{\pm}\right]=\pm\hat{K}_{\pm}$ and $\left[\hat{K}_{+},\hat{K}_{-}\right]=2\hat{K}_{0}$.  The time evolved density operator is expressed as
\begin{gather}\hat{\rho}(\tau)=\frac{1}{z_{0}}\exp{\left(\beta\omega_{0}\left(1-a(\tau)2\hat{K}_{0}-c(\tau)\hat{K}_{-}-c(\tau)^{*}\hat{K}_{+}\right)\right)},\end{gather}
where $a(\tau)$ is given in Eq. \eqref{eq:atau} and 
\begin{eqnarray}
c(\tau)=-\frac{g}{\Omega}(1+2|v(\tau)|^{2})+\nonumber \\
+2i\mathrm{Im}(u(\tau)v(\tau)^{*})-\frac{2\omega(\tau)}{\Omega}\mathrm{Re}(u(\tau)v(\tau)^{*})
\label{eq:c2}
\end{eqnarray}
where $u(\tau)$ and $v(\tau)$ are the Bogoliubov coefficients defined in Eq. \eqref{eq:btime}). In Eqs. \eqref{eq:atau} and \eqref{eq:c2}, $\omega(\tau)$ is the non-interacting energy with renormalized velocity, $g$ is the interaction strength and $\Omega=\sqrt{\omega(\tau)^{2}-g^{2}}$ is the eigenenergy of the final 
Hamiltonian. \jav{For arbitrary quench protocol, $a(\tau)^{2}-|c(\tau)|^{2}=1$ holds true and, hence, $a(\tau)\geq 1$.}
The other exponential under the trace in Eq. \eqref{eq:fdef2} is rewritten as
\begin{gather}e^{i(\xi_{+}\hat{n}_{+}+\xi_{-}\hat{n}_{-})}=e^{-i\xi+i\Delta\xi\Delta\hat{n}}e^{i\xi\,2\hat{K}_{0}},\end{gather}
where we have introduced $\xi=(\xi_{+}+\xi_{-})/2$ and $\Delta\xi=(\xi_{+}-\xi_{-})/2$. The operator $\Delta\hat{n}=\hat{n}_{+}-\hat{n}_{-}$ commutes with both $\hat{K}_{0}$ and $\hat{K}_{\pm}$.
Using a faithful representation of $su(1,1)$ algebra \cite{solomon,gilmore}, we derive a single exponential which equals the product $\hat{\rho}(\tau)e^{i(\xi_{+}\hat{n}_{+}+\xi_{-}\hat{n}_{-})}$. The generators of the $su(1,1)$ algebra may be faithfully represented by $2\hat{K}_{0}\rightarrow \sigma_{z}$ and 
$\hat{K}_{\pm}\rightarrow  (\pm\sigma_{x}+i\sigma_{y})/2$ where $\sigma_{x}$, $\sigma_{y}$ and $\sigma_{z}$ are the $2\times 2$ Pauli matrices. The single exponent then can be diagonalized by standard Bogoliubov transformation. After diagonalization,
\begin{gather}
\hat{\rho}(\tau)e^{i(\xi_{+}\hat{n}_{+}+\xi_{-}\hat{n}_{-})}= \frac{e^{-i\xi+\beta\omega_{0}}}{z_{0}} \nonumber \\
\times \exp{\left(i\Delta\xi\Delta\hat{\bar{n}}-\ln(B+\sqrt{B^{2}-1})(1+\hat{\bar{n}}_{+}+\hat{\bar{n}}_{-})\right)},
\label{eq:sexp}
\end{gather}
where
\begin{gather}B=\cos\xi\cosh(\beta\omega_{0})-ia(\tau)\sin\xi\sinh(\beta\omega_{0}),\end{gather}
and $\hat{\bar{n}}_{\pm}=\bar{d}^{+}_{\pm}\bar{d}_{\pm}$ is the occupation number operator after Bogoliubov transformation and $\Delta \hat{\bar{n}}=\hat{\bar{n}}_{+}-\hat{\bar{n}}_{-}$. The annihilation operator is expressed with the  new annihilation and creation operators as
\begin{gather}
d_{\pm}=\frac{\bar{d}_{\pm}+\gamma\bar{d}^{+}_{\mp}}{\sqrt{1-|\gamma|^{2}}}, \\
\gamma=\frac{\sqrt{B^{2}-1}+i\sin\xi\cosh(\beta\omega_{0})-a(\tau)\cos\xi\sinh(\beta\omega_{0})}{c(\tau)e^{-i\xi}\sinh(\beta\omega_{0})}.
\end{gather}
Substituting Eq. \eqref{eq:sexp} into Eq. \eqref{eq:fdef2}, the generating function is obtained as
\begin{eqnarray}
f(\xi_{+},\xi_{-})=\Big[1+n(\tau)\left(1-e^{i(\xi_{+}+\xi_{-})}\right)+ \nonumber \\
+(n_{0}+n_{0}^{2})\left(e^{i\xi_{+}}-1\right)\left(e^{i\xi_{-}}-1\right)\Big]^{-1}
\end{eqnarray}
where $n_{0}=(e^{\beta\omega_{0}}-1)^{-1}$ and $n(\tau)=\mathrm{Tr}\left[\hat{\rho}(\tau)\hat{n}_{\pm}\right]=\left(a(\tau)(2n_{0}+1)-1\right)/2$ are the expectation value of the occupation number before and after the quench.
Note that $f(\xi_{+},\xi_{-})$ is a $2\pi$-periodic function of its variables and $f(\xi_{+},\xi_{-})=f(\xi_{-},\xi_{+})$. The latter implies $\rho(n_{+},n_{-})=\rho(n_{-},n_{+})$.

\section{Fidelity}
\label{sec:fid}
{\jav{The technical difficulty in computing the fidelity is evaluating the trace of some exponentials, the exponent of which are expressed in terms of $\hat{K}_{0}$, $\hat{K}_{\pm}$ and $\Delta \hat{n}$, irrespectively of the norm chosen on the set of density operators.}} Using again the faithful representation of the $su(1,1)$ algebra, the product of exponentials can be transformed into a single exponential in the same way as in Appendix \ref{sec:Fxi}. In order to calculate the square root of an exponential, which we need in the case of the Uhlmann fidelity, we diagonalize the exponent and halve the eigenvalues. The trace of the single exponential is evaluated after diagonalizing the exponent.
Using this procedure, we obtain the fidelity using both the Frobenius and the Bures metric\cite{zanardi}. 

The Uhlmann fidelity is evaluated as
\begin{gather}\ln F_{U}(t)=\ln\mathrm{Tr}\left[\sqrt{\hat{\rho}^{1/2}_{0}\hat{\rho}(t)\hat{\rho}^{1/2}_{0}}\right]=\nonumber \\
=\sum_{q>0}\ln\frac{\cosh(\beta\omega_{0}(q))-1}{\sqrt{1+|u_{q}(t)|^{2}\sinh^{2}(\beta\omega_{0}(q))}-1}\,.\end{gather} 
In Eq. \eqref{eq:fff}, $u_{q}(t)$ is the Bogoliubov coefficient defined in Eq. \eqref{eq:btime}. 
Up to second order in $g_{2}/v$ and for a SQ,
\begin{gather}|u_{q}(t)|\approx 1+\frac{g(q)^{2}}{2\omega_{0}(q)^{2}}\sin^{2}(\omega_{0}(q)t),\end{gather} and
\begin{widetext}
\begin{gather}\ln F_{U}(t)=-{\displaystyle\sum_{q>0}}\dfrac{\frac{g^{2}}{\omega_{0}^{2}}\sin^{2}(\omega_{0}t)}{1+\tanh^{2}\left(\frac{\beta\omega_{0}}{2}\right)}=\nonumber \\
=-\alpha\left[1+\dfrac{\tau_{0}}{\beta}\left(\psi\left(\dfrac{3}{4}+\dfrac{\tau_{0}}{2\beta}\right)-\psi\left(\dfrac{1}{4}+\dfrac{\tau_{0}}{2\beta}\right)\right)+\mathrm{Re}\left(\frac{i\tau_{0}}{t-i\tau_{0}}-\frac{\tau_{0}}{\beta}\left(\psi\left(\dfrac{3}{4}+\dfrac{\tau_{0}}{2\beta}-\frac{it}{2\beta}\right)-\psi\left(\dfrac{1}{4}+\dfrac{\tau_{0}}{2\beta}-\frac{it}{2\beta}\right) \right)\right)\right]
\label{eq:fut}
\end{gather}
\end{widetext}
where $\psi(x)$ is the digamma function.
 In the long time limit the last term in Eq. \eqref{eq:fut} converges to zero.
At low temperatures,
\begin{gather}\ln F_{U}(t\gg\beta\gg\tau_{0})=-\alpha\big(1+\pi\tau_{0}T\big)\end{gather}
where we used $\psi(3/4)-\psi(1/4)=\pi$.

Using the Frobenius norm, the overlap of the time-evolved and initial states is derived as
\begin{gather}\ln F_{F}(t)=\frac{1}{2}\ln\mathrm{Tr}\left[\hat{\rho}(t>\tau)\hat{\rho}_{0}\right]=\nonumber \\
=\sum_{q>0}\ln\frac{\cosh(\beta\omega_{0}(q))-1}{|u_{q}(t)|\sinh(\beta\omega_{0}(q))}\,.\end{gather}
The effective dimension,
\begin{gather}\ln d_{\mathrm{eff}}=-\ln \mathrm{Tr}\left[\rho_{0}^{2}\right]=2\ln Z_{0}(\beta)-\ln Z_{0}(2\beta)=\nonumber \\
=2\sum_{q>0}\ln\frac{1-e^{-2\beta\omega_{0}(q)}}{(1-e^{-\beta\omega_{0}(q)})^{2}}\end{gather}
leads to the temperature independent normalized fidelity
\begin{gather}\ln \left(\sqrt{d_{\mathrm{eff}}}F_{F}(t)\right)=-\sum_{q>0}\ln|u_{q}(t)|\label{eq:fff}\end{gather}
for arbitrary quench protocol.  Within perturbation theory and for a SQ, we obtain
\begin{gather}\ln \left(\sqrt{d_{\mathrm{eff}}}F_{F}(t)\right)=
-\sum_{q>0}\frac{g(q)^{2}}{2\omega_{0}(q)^{2}}\sin^{2}(\omega_{0}(q)t)=\nonumber\\ =-\alpha\frac{t^{2}}{\tau_{0}^{2}+t^{2}}.
\end{gather}
 For large times ($t\gg\tau_{0}$) the normalized fidelity saturates at $e^{-\alpha}$.

\end{document}